\documentclass[11pt]{article}
\usepackage{amsmath}
\usepackage{lipsum} 
\usepackage[margin=2.54cm]{geometry}
\usepackage{geometry}
\usepackage[utf8]{inputenc}
\usepackage{xcolor}
\usepackage{booktabs}
\usepackage{setspace}
\usepackage{graphicx}
\usepackage[hidelinks]{hyperref}
\usepackage[document]{ragged2e}
\usepackage{pythonhighlight}
\usepackage{pdflscape}
\usepackage{longtable}
\usepackage{epstopdf}
\usepackage[flushleft]{threeparttable}
\usepackage{comment}
\usepackage{tikz}
\usepackage{standalone}
\usepackage{array}
\usepackage{adjustbox}
\usepackage{svg}
\usepackage{dutchcal}
\usepackage{changepage}
\usepackage{dutchcal}
\usepackage{caption}
\usepackage{subcaption}
\usepackage{soul}
\usepackage{float}
\usepackage{mdframed}
\usetikzlibrary{mindmap,trees}
\usepackage{stackengine}
\usepackage{afterpage}
\usepackage{natbib}
\usepackage[affil-it]{authblk}

\definecolor{quotemark}{gray}{0.7}
\makeatletter
\def\fquote{%
    \@ifnextchar[{\fquote@i}{\fquote@i[]}%]
           }%
\def\fquote@i[#1]{%
    \def\tempa{#1}%
    \@ifnextchar[{\fquote@ii}{\fquote@ii[]}%]
                 }%
\def\fquote@ii[#1]{%
    \def\tempb{#1}%
    \@ifnextchar[{\fquote@iii}{\fquote@iii[]}%]
                      }%
\def\fquote@iii[#1]{%
    \def\tempc{#1}%
    \vspace{1em}%
    \noindent%
    \begin{list}{}{%
         \setlength{\leftmargin}{0.1\textwidth}%
         \setlength{\rightmargin}{0.1\textwidth}%
                  }%
         \item[]%
         \begin{picture}(0,0)%
         \put(-15,-5){\makebox(0,0){\scalebox{3}{\textcolor{quotemark}{``}}}}%
         \end{picture}%
         \begingroup\itshape}%

 \def\endfquote{%
 \endgroup\par%
 \makebox[0pt][l]{%
 \hspace{0.8\textwidth}%
 \begin{picture}(0,0)(0,0)%
 \put(15,15){\makebox(0,0){%
 \scalebox{3}{\color{quotemark}''}}}%
 \end{picture}}%
 \ifx\tempa\empty%
 \else%
    \ifx\tempc\empty%
       \hfill\rule{100pt}{0.5pt}\\\mbox{}\hfill\tempa,\ \emph{\tempb}%
   \else%
       \hfill\rule{100pt}{0.5pt}\\\mbox{}\hfill\tempa,\ \emph{\tempb},\ \tempc%
   \fi\fi\par%
   \vspace{0.5em}%
 \end{list}%
 }%
 \makeatother

\newmdenv[linecolor=black,linewidth=1pt,roundcorner=10pt]{hypothesis}

\usepackage{rotating}

\newcolumntype{R}[2]{%
    >{\adjustbox{angle=#1,lap=\width-(#2)}\bgroup}%
    l%
    <{\egroup}%
}
\renewcommand{\arraystretch}{1.5}

\title{Drivers and Barriers of AI Adoption and Use in Scientific Research\footnote{The research leading to the results of this paper has received financial support from the French National Research Agency [reference: SEED -ANR-22-CE26-0013-01].}} %We also thank the fruitful exchanges with colleagues in DG-RTD European Commission, especially: David Arranz, Valentina Di Girolamo, Daniela Petkova, Julien Ravet, and Laura Roman.}}

\author[1]{Stefano Bianchini}
\author[1]{Moritz M\"uller}
\author[1,2  \footnote{Email: \texttt{s.bianchini@unistra.fr} ; \texttt{mueller@unistra.fr} ; \texttt{p.pelletier@unistra.fr}}
]{Pierre Pelletier}

\affil[1]{\small BETA -- University Strasbourg, France}
\affil[2]{\small CPS -- University of Turin, Italy}

\date{}
\begin{document}
\doublespacing
\justifying

\maketitle

\begin{abstract}
New technologies have the power to revolutionize science. It has happened in the past and is happening again with the emergence of new computational tools, such as artificial intelligence and machine learning. Despite the documented impact of these technologies, there remains a significant gap in understanding the process of their adoption within the scientific community. In this paper, we draw on theories of scientific and technical human capital to study the integration of AI in scientific research, focusing on the human capital of scientists and the external resources available within their network of collaborators and institutions. We validate our hypotheses on a large sample of publications from OpenAlex, covering all sciences from 1980 to 2020, and identify a set key drivers and inhibitors of AI adoption and use in science. Our results suggest that AI is pioneered by domain scientists with a `taste for exploration' and who are embedded in a network rich of computer scientists, experienced AI scientists and early-career researchers; they come from institutions with high citation impact and a relatively strong publication history on AI. The access to computing resources only matters for a few scientific disciplines, such as chemistry and medical sciences. Once AI is integrated into research, most adoption factors continue to influence its subsequent reuse. Implications for the organization and management of science in the evolving era of AI-driven discovery are discussed.
\end{abstract}

%\keywords{Artificial intelligence; Technology adoption; Organization and management of science}

\clearpage

%\thispagestyle{empty}
%\tableofcontents

%\thispagestyle{empty}

\section{Introduction}
\label{sec:introduction}

The pace of scientific progress is a direct correlate of our alliance with new technologies. It has been so in the past with instruments such as the microscope and telescope, more recently with computers and the internet \citep{rosenberg1992, ding2010}. Today the spotlight is on AI/ML, which is fast emerging in the scientific landscape as a tool with great potential for discovery, a new ``general method of invention'', as many have recently argued \citep{agrawal2018finding, cockburn_2018, furman2020,  bianchini2022artificial, hain2023}.\footnote{\justifying For simplicity, throughout the article, we will use the appellation ``AI'' instead of ``AI/ML''.} Problems that once were impossible to contemplate, or even formulate, come around every day. Examples include predicting the 3D structure of proteins \citep{jumper2021highly}, regulating nuclear fusion plasma in the tokamak configuration \citep{degrave2022magnetic}, predicting the formation of the structure of the universe \citep{he2019learning}, and creating a map of the brains of small insects \citep{winding2023connectome}. The potential of AI to accelerate and advance scientific discovery is being explored across almost all scientific disciplines, and at different stages of research processes (see, e.g., \citealp{wang2023nature}). 

%The potential of AI to accelerate and advance scientific discovery is being explored across almost all scientific disciplines, and at different stages of research processes (see, e.g., \citealp{wang2023nature}). But as the potential of AI in science grows, it becomes essential to understand what resources are critical for scientists to successfully adopt this technology, unleash its impact, and facilitate its democratization throughout the scientific system, ensuring that no one is left behind.

As the potential of AI in science grows, it becomes essential to understand what resources are critical for scientists to successfully adopt this technology throughout the science system. The paper at hand addresses this issue by comparing the resource endowment of domain scientists who have (persistently) adopted AI versus those who have not.

A recent \textit{Nature} survey \citep{VannoordenPerkel2023} of more than 1,600 researchers worldwide asked scientists who have already experimented with AI what obstacles prevent them from using it more and received the following responses: `Lack of skills or skilled researchers' (80\%), `Lack of training resource' (70\%), `Lack of funding' (50\%), Lack of computing resources (35\%), `Lack of data' (30\%), and others (18\%). In contrast, scientists who have no experience with AI mostly declared that they do not find it useful for their research.  It thus seems clear that AI adoption is inextricably linked, at a very least, to an insufficient understanding of the technology's potential in a specific domain, individual skills, and the accessibility of computing capacity and data (also confirmed in, e.g., \citealp{hwang2018, thompson2022}). The bibliographic study by \citet{ThuEtAl2022}, on the composition of scientific teams in applied AI, shows that the lack of skills on the part of domain scientists is often compensated for by engagement in interdisciplinary research with computer scientists, and that such interdisciplinary efforts are also successful in terms of citations received. However, not all scientists may be in a position to implement, or even consider, such a collaborative strategy. What (other) resources then favor AI adoption? Also, not all scientists who try AI intend to adopt it persistently in their research -- in our sample that share is about 50\%. What are the barriers that prevent these scientists from integrating AI into their research?\footnote{The study of AI diffusion and, especially, adoption has hitherto focused on firms and not, to our knowledge, on scientists. In recent years, research has examined the various technological, organizational and environmental prerequisites for the (successful) adoption and use of AI technologies in companies, considering factors such as company size, industry sector and digital skills (see, e.g., \citealp{bughin2017, alsheibani2018, enholm2022, kinkel2022, mcelheran2024}). This strand of literature has been helpful for our conceptual framework because several factors important to business find close parallels in the context of science -- e.g., the size of a research team or the peculiarities of a scientific domain.}

Particularly apposite to our research question is the theory of scientific and technical human capital (STHC) because it captures the idea that scientists' behavior is contingent not only on their own human capital -- various kinds of knowledge and skills that are `internal' to the individual -- but also on a larger reservoir of resources that reside in their working relations \citep{bozeman2001scientific, bozeman2004scientists}. Scientists do not exist in a social vacuum and the production of scientific knowledge is inherently a social enterprise, just like, as we argue in this paper, the integration of new technologies into scientific practices. 
Here, we consider three dimensions of STHC that can influence the decision of scientists to adopt AI in their respective domain: (i) their personal pre-existing knowledge, skills, and taste for experimentation; (ii) the knowledge and expertise of their research collaborators; and (iii) the institutional setting in which the researcher is embedded. Social ties are relevant in that, through collaborations, scientists can acquire and employ complementary skills and technical resources to create and transform knowledge and ideas in ways that would not be possible in an isolated context \citep{bozeman2004scientists,taylor2006superman,lee2015creativity, leahey2016sole}. The institutional setting is another important factor because it shapes the practice of ‘‘doing science'' within an organization -- from running physical infrastructure and raising funds to ethical norms of scientific conduct -- and ultimately molds the research trajectories of its members \citep{fox1991gender,heinze2009organizational,fortunato2018science}. 

Scientific knowledge and expertise will be measured on past publications, assessing thematic diversity, impact, collaborations with peers and other dimensions; the institutional setting will encompass the quality of the researcher's home institution, reputation, research orientation, and access to high-performance computing (HPC), presumably crucial for modern AI research. As for the econometric strategy, we will model AI adoption in research papers as a function of technology available in a given field at a particular time and various aspects of scientists' STHC. Our approach, based on conditional logit regression with matched pairs (adopters \textit{vis-à-vis} non-adopters), allows us to remove other unobservable components -- i.e., cohort and field interactions with dynamic AI technologies -- that might influence the adoption decision. 

% I WOULD NOT STRESS LIMITATIONS HERE IN THE INTRO - MOVED TO SECTION RESUTS Our main findings are detailed in the abstract. Interpretation of coefficient estimates must be cautious for mainly two reasons. Firstly, our regressions take into account the dynamic nature and variety of the technology but do not allow to identify one-way causal relationships. In particular, we find that AI adopters tend to be embedded in local environments that are rich with AI relevant resources, but that may be partly due to reverse causality (scientists that are eager to adopt form these environment), as well as due to self-selection (scientists with certain unobserved skills or tastes may by more likely to adopt AI as well as end up in preferable environments). We also note that some measures may capture a variety of factors. Consider for example scientific impact of the institution measured by the stock of citations received; this measure may proxy a wealth of human capital in the institution but also proxy physical infrastructure or norms.

% Si jamais l'anciene version intro est à la fin du fichier TODO.tex

Among the various findings from this study, we believe the following are noteworthy. First, AI adoption is mainly shaped by social factors. Indeed, previous collaborations with computer scientists and AI experts strongly predict adoption, as does affiliation with institutions specializing in AI research. The relevance of prior ties with individuals with strong AI know-how suggests that the adoption process occurs at the boundary, or intersection, of computer science and AI application domains, and that the transfer of knowledge from development (in computer science) to application (in another domain) hinges on inter-disciplinary research collaborations. We contend that these collaborations are essential to offer valuable ``about knowledge'' on AI -- i.e., fairly simple facts and information about the potential of technology -- enabling scientists to grasp the possibilities of integrating new tools into their respective domains. Second, while domain scientists benefit from social ties with individuals knowledgeable in AI, an excessive number of computer scientists participating in the initial phases of AI adoption may hinder subsequent re-use. This is most likely due to an excessive dependence on specialized computing skills that do not readily transfer to domain scientists. It also suggests that the current state of AI technology, particularly software tool-kits, may not be sufficiently adequate to streamline the implementation of AI techniques for non-AI experts. Third, early-career researchers play a pivotal role in the adoption process, challenging the conventional view that mentors solely impart knowledge and shape the research trajectory to young scientists. Our findings suggest that the younger generation brings the necessary skills required in the era of AI-driven scientific discovery and may remold the research agenda of their senior peers. Finally, despite the fact that modern AI relies heavily on computing power, we find that the local availability of advanced computing infrastructure correlates positively with AI adoption in only a few areas, namely medical sciences and chemistry. These findings, along with others, have significant implications for the organization and management of science, which will be discussed in more detail in the final section of this manuscript. \\

\section{STHC endowment and AI adoption}

In this section, we apply a revised version of the STHC framework to the context of new technology adoption in science and establish a set of testable hypotheses. We begin by exploring the influence of social relations and network ties among scientists, as well as the role of the institutional environment in which they operate, including access to computational resources. We then discuss the role of internal resources, such as formal education and past experience, along with other individual traits of the researcher. In articulating our hypotheses, we provide some information on the measurement of variables.

\subsection{External resources}

\subsubsection{Social capital and about-knowledge from social ties}

% Stefano - Re-touch this if we want to keep it.  
%Existing social connections may play a significant role in AI adoption through at least two channels: becoming part of a productive team dealing with AI and the ability to judge potential relevance of AI for one's own research.

%Social ties offer a form of social capital that can be advantageous for researchers when adopting AI, as they might directly contribute to the initial AI project. As Bozeman (2001) emphasized, \textit{[A]t the project S\&T human capital level, the focus is on the aggregate of all project participants' endowments and social connections, as well as the physical and economic resources available to a project} \citep[p.20]{bozeman2001scientific}. For instance, an applied chemist may collaborate with computer scientists to use AI methodologies in a joint research endeavor. In this case, collaborating computer scientists could be either previous collaborators or discovered through an existing social network. Either way, having prior collaborations with computer scientists may be helpful in establishing collaborations with them in the (future) AI project.

Past interactions and collaborations with peers influence a scientist's perceptions and interpretations of novel technological advancements such as AI. But to embark on a new field, what knowledge should a scientist possess? The prevailing belief is that for a scientist who wants to venture into a new field or incorporate methodological tools from that field into their own research, it is desirable to ‘know more' about \textit{what is going on} in the field. While we do not dismiss that knowing more is desirable (albeit one may wonder in what amount), we believe it is not the only prescription for a scientist to build bridges between domains. Nor is it perhaps the most efficient prescription. Science is simply too big and, as a result, there are cognitive limits that prevent individuals from fully understanding disciplinary tools, knowledge architectures, and associated ‘best practices' across its various domains \citep{jones_2009, park2023}. This is especially true for emerging technologies, whose potential and applications in science are not yet entirely clear \citep{rotolo2015emerging}. What specifically needs to be known then? And for what purposes?

The theory of ‘about-knowledge' offers a compelling answer. It suggests the existence of a specific type of knowledge, namely \textit{about-knowledge} or \textit{connective knowledge}, which helps scientists recognize the potential that could be realized by merging their own expertise with knowledge from other fields -- in our case, realize the potential of AI as scientific tool in a given application domain.

About-knowledge can be thought as the kind of ‘know-how' that is necessary to achieve an early point of connection with another field -- ``\textit{[A] range of fairly simple facts and information about the sort of problem domains and approaches that populate different fields and specialisms}'' \citep[p. 8]{priaulx2018connective}. It is not an intimate understanding of how the scientist's core domain expertise interacts with another field, but rather a set of cognitive foundations into the kind of contributions that the field can make to their own expertise, as well as a broad understanding of the practical settings, languages, sub-cultures, expectations and reward models that regulate that field.%\footnote{\justifying It is noteworthy that this type of knowledge and resulting connections, which we believe can be immensely valuable, particularly in the initial stages of the collaborative life cycle, are often overlooked in the literature on interdisciplinarity and transdisciplinarity.}  

It is clear that the lack of insights into what AI scientists do and/or where AI research is headed significantly decreases the likelihood that a scientist will recognize the relevance of the field's contribution to their own work. Recent research has shown that the same dynamics apply to AI adoption and diffusion within companies: uncertainty about the technology's use cases (i.e., what to do with it and what competitive advantages it can bring) slows down adoption \citep{ameye2023}. In this sense, about-knowledge should fill the deficit by providing a wide-angled lens of the potential of the technology, thus prompting the scientist to give it a try. 

One may argue that about-knowledge is not knowledge at all. Instead, it is simply a series of decontextualized facts or accounts of popular understanding. But this is precisely what makes the concept of about-knowledge so important to our research, as our hypothesis is that the ‘big picture' about AI acts as an initiating force that motivates scientists to pursue the idea of incorporating it into their research. How do researchers acquire about-knowledge related to AI? How does this knowledge cross disciplinary boundaries? And how can we eventually operationalize the concept of about-knowledge in practice?

In recent years numerous initiatives have encouraged the next generation of scientists to pursue interdisciplinary programs with a focus on AI and data science more generally. These efforts are rooted in the conviction that formal education and training are necessary to close the knowledge gap in this field – which may be partially true, although discussing in depth the true benefits of interdisciplinarity training is beyond the scope of our research. The mechanisms of knowledge transmission, especially about-knowledge, however, may be much simpler. We will simply contend that even a little insight into other fields can help the scientist understand the role that these fields can play in their work and, as in the context of AI, help see the potential of the technology and address misconceptions and non-conceptions that would otherwise remain mistakenly overlooked. 
 
And here is where the STHC framework finds a link with the theory of about-knowledge: it is about people embedded within the collaboration networks and populating the same institutional environment who can facilitate knowledge exchange and mediate interactions. Social ties are simply the most critical vehicle for enhancing about-knowledge connectivity across diverse domains.

So let us start from collaboration networks. Here, we use the network of past collaborations of domain scientists as a valuable historical record of their interactions and collaborations with peers (see, e.g., \citealp{katz1997, newman2001, wang2017}), particularly those with some experience in AI technology within the same application domain or in computer science more in general. It is reasonable to assume, in fact, that scientists who are embedded in a network where their peers have already proven experience with a certain technology (e.g., as evidenced by at least one publication) have more incentives to adopt the same technology in their own research, especially if their colleagues have achieved successful outcomes. This is because, even if a collaboration does not occur with a past collaborator, being part of a network reinforces a scientist's rudimentary knowledge about the potential of the technology. More formally, we argue that: \\

\begin{hypothesis}
\noindent \textbf{H1a}: \textit{Prior ties to scientists with AI-relevant human capital increase the likelihood that a domain scientist adopts AI.} 
\end{hypothesis}

\vspace{1em}

The second channel through which knowledge and about-knowledge can reach a domain scientist is through social interactions with their peers who work in the same institution. Organization science has long established that the location of an actor's contacts in the social structure can offer advantages to the actor when it comes to acquiring information and resources, as do attributes that are rooted in their interactions, such as trust and trustworthiness \citep{tsai1998social}. This is because communication is a complex and often arduous process that requires individuals to converge on a common sense and is thus facilitated by both spatial and cognitive proximity.\footnote{\justifying It should be noted that in this context, ‘cognitive' refers more to the understanding of collective objectives that are shared by a group of individuals or an organization (see, e.g., \cite{coleman1988social}), rather than the similarities of knowledge bases between individuals. The literature suggests that reciprocity (i.e., a favor for a favor; an action for an action) and a sense of contribution to the organization are two key factors that encourage knowledge and information sharing between individuals within an organization \citep{cummings2004work,wang2010knowledge}.}

It is plausible to assume that individuals who work in institutions that specialize in AI research have easier access to knowledge and about-knowledge pertaining to AI. On the one hand, members of the same institution have more opportunities to spend time together on social occasions and hence more opportunities to exchange ideas and resources freely -- i.e., through casual conversations and informal communications instead of formalized meetings and events. On the other hand, members of an institution working in close (geographical and cognitive) proximity may exhibit mimetic isomorphic behaviors, whereby they tend to adopt similar structures, practices, and strategies to their peers, a concept first described to explain what makes organizations so similar \citep{dimaggio1983iron, mizruchi1999social}. We believe that this phenomenon can also be observed in scientific research, where researchers may adopt similar research designs, methods, and theoretical frameworks to those used by their colleagues within the same institution, particularly when facing uncertainty or ambiguity, as we can assume in the context of AI adoption. Taken together the above arguments lead us to the following research hypothesis: \\ 

\begin{hypothesis}
\noindent \textbf{H1b}: \textit{A prevalence of AI research within an institution increases the likelihood that a domain scientist adopts AI.} 
\end{hypothesis}

\vspace{1em}

\subsubsection{Mentorship and newbies}

What is more important than the mentor-newbie relationship when it comes to social ties and knowledge transfer? The term ‘mentor' typically denotes an experienced individual who imparts their skills and knowledge to a younger person, often someone identified as promising and part of the next generation (e.g., post-doctoral researchers, PhD students, or junior untenured researcher) \citep{archibugi2021choosing}. Under the right circumstances, a mentorship collaboration can facilitate the transfer of various S\&T human capital assets, such as craft-skills, know-how, contacts with other peers, industry and funding agents, and more. However, in the context of new methods for scientific discovery and fresh ways to approach scientific problems, it is reasonable to assume that the flow of assets could also occur in the reverse direction, that is \textit{from the junior to the mentor}. And we have strong evidence to support this conjecture.

The academic job market is rich in human resources specializing in AI. According to some recent statistics, the number of AI-related curricula has increased more than any other curriculum in recent years and is unlikely to slow down in the years to come. For instance, in 2020 alone, over 30,000 undergraduate students in the US completed a computer science degree, and one in every five students who earned a PhD degree in computer science specialized in AI.\footnote{\justifying See Stanford AI Index Report from 2022 here: \href{https://aiindex.stanford.edu/report/}{https://aiindex.stanford.edu/report/}} We can expect similar figures in many other countries. AI-related courses are no longer limited to computer science departments at the undergraduate level; rather, a growing number of universities offer interdisciplinary programs that combine AI with other fields. The new generation of scientists also has at their disposal a plethora of online resources offered by universities and private companies that focus specifically on AI. One example is Massive Open Online Courses (MOOCs), which are emerging as an affordable and popular option for those who want to deepen their knowledge of AI, from introductory courses to others on cutting-edge algorithms and advanced applications.%\footnote{\justifying The importance of AI literacy from the early stage of education has also been recognized globally. A recent report by UNESCO (2022) highlights the commitment of several countries to developing AI literacy and competencies in K-12 schools. Generally, these initiatives aim to prepare new generations for a world in which AI will be ubiquitous, and thus understand the power and versatility of this technology along with its ethical dilemmas.}  

In summary, we are confident that young researchers who are well-versed with AI techniques and tools can bring new perspectives and insights to more experienced colleagues who may be stuck in doing science ``as usual''. Empirically, we will identify newbies as authors who have published for the first time in a given year. We posit that: \\

\begin{hypothesis}
\noindent \textbf{H2}: \textit{Prior ties to early-career researchers increases the likelihood that a domain scientist adopts AI.} 
\end{hypothesis}

\vspace{1em}

\subsubsection{Computational resources}

While AI is commonly perceived as an intangible technical system, it is \textit{de facto} rooted in physical infrastructure and hardware. Yet, the role of physical assets and their associated computing capabilities -- also known as AI compute -- have been largely overlooked in policy circles and scholarly literature.%\footnote{\justifying One reason for that is the lack of standardized and validated data on computing resources. National and institutional and data on the supply and demand of AI compute is not easily accessible and, in some cases, considered sensitive proprietary information.} 

AI compute can be understood as ``\textit{one or more stacks of hardware and software used to support AI workloads and applications in an efficient manner}''\citep[p.20]{/content/paper/876367e3-en}. For machine learning systems, it is clear that compute can facilitate three key steps in the scientific pipeline: i) processing and cleaning large data, ii) training models and calibrating them (e.g., determining the value of weights of a neural network from the data presented to the model), and iii) inferencing, which is using the trained model for a specific application to determine an output. Of course, the computing requirements can vary considerably depending on the application, ranging from large high-performance computing (HPC) clusters to smaller laptops and workstations.

Cutting-edge research in AI has become synonymous with access to large computing infrastructures and expertise to exploit them. \cite{sevilla2022compute}, for example, carried out a detailed investigation of the computational requirement of 123 milestone ML models over time and showed that since the 2010s, the amount of computation required to accommodate modern machine learning systems has soared, with an impressive 5.7-month doubling time (see also \cite{AmodeiHernandez2018} for estimates with different assumptions) -- just for comparison, remember that Moore's law has a 2-year doubling period. While not all researchers use state-of-the-art and computationally intensive ML systems, having access to computing resources can still make a significant difference and, reasonably, be a major driver of AI adoption. How then can scientists access computing resources?

Researchers have various options for accessing AI compute, including data centers or supercomputers located in physical facilities, public or private cloud computing services, and decentralized access at the edge of devices, such as mobile IoT devices. It can be difficult to empirically determine which resource(s) a researcher relies on for their work, yet we contend that the local availability of computing resources may serve as a motivating factor for researchers to adopt AI for the first time and potentially use it again. This is not just because researchers can handle larger and more complex datasets and get results faster than they could with limited computing resources, but also because of the institutional culture that embraces AI, as we discussed in Section 2.2.1. Scientists are well-aware that computing resources are readily available and can potentially support their work; they also know that they can rely on support services to help them use these resources more efficiently and effectively.

Here, we measure the availability of AI compute by the presence of an HPC cluster within the focal researcher's organization (although it should be noted that such compute infrastructure can also be used for non-AI workloads such as mathematical modeling and simulations). In very general terms, HPC is a technology that uses clusters of powerful processors, working in parallel, to process data and solve complex problems at high speeds \citep{/content/paper/876367e3-en}. Unlike standard computing systems, HPC systems can handle multiple tasks simultaneously across multiple computer servers or processors with a centralized scheduler that manages the computing workload. The high cost of HPC can put this technology out of reach for most organizations, resulting in a significant ``compute divide'' within and between countries and institutions, as well as between the private sector and academia \citep{ahmed2020democratization, ahmed2023}. This is especially true for AI applications in some data-intensive scientific fields such as genomics and bioinformatics or particle physics where ML training and inferencing can be highly demanding in terms of memory and computational resources. The existence of a computational divide can therefore impede the adoption of AI and generate disparities in the productivity gains that AI can offer to science. 

In summary, our hypothesis is that: \\

\begin{hypothesis}
\noindent \textbf{H3}: \textit{The presence of HPC cluster within a researcher's organization increases the likelihood of adopting AI.} 
\end{hypothesis}

\vspace{1em}

\subsection{Internal resources}

We now turn to the `internal' resources of the domain scientist, which can be broadly classified into three, somewhat overlapping, categories: i) cognitive skills, ii) scientific and technical knowledge, and iii) contextual skills \citep{bozeman2001scientific}.

\subsubsection{Scientific background and experience}

Let us start with the most straightforward, namely scientific and technical knowledge. This is the type of knowledge acquired through formal scientific education; it involves a thorough understanding of particular theories, experimental and research findings, and the ability to anticipate where research in a particular area is heading. From a Kuhnian perspective, scientific and technical knowledge enables the scientist to feel part of a specific epistemic community, to be accepted by their peers as a member of that community, and ultimately to adopt the shared scientific paradigm \citep[Ch.2 and 3]{kuhn1962structure}. 

Contextual skills can be viewed instead as a subset of scientific and technical knowledge and relate more closely to the type of knowledge gained from practical research experience. Unlike scientific and technical knowledge, contextual skills often involve a tacit component that can only be obtained “on-the-job”, that is, in the process of doing research. 

In our study, we use a scientist's first field of activity (i.e., domain of the first publication) as a proxy for formal scientific education and context skills. While we do not advance any specific research hypotheses, we believe that this variable is essential to account for idiosyncratic differences in the propensity to adopt AI that may arise due to an individual's scientific background.

\subsubsection{Taste for exploration}

The third dimension is about cognitive skills, that can be viewed as those mental abilities and processes that allow individuals to perceive, process, and use information in a given environment. As such, they necessarily relate to science, but not exclusively to it; they include skills such as reasoning, learning, and others. Here, we are particularly interested in the dispositions or traits that underlie many cognitive skills and processes. One of these is a \textit{taste for exploration}. 

Exploration is intimately linked to curiosity, a personal trait that prompts individuals to explore uncharted territories. We think that curiosity is a useful construct for understanding scientists' behavior in terms of technology adoption. Although psychologists have not reached a consensus on its definition, it is generally accepted that curiosity involves an intrinsic motivated desire for new information – an ``appetite for knowledge'', or more formally ``\textit{a form of cognitively induced deprivation that arises from the perception of a gap in knowledge or understanding}'' \citep[p. 75]{loewenstein1994psychology}.%\footnote{\justifying  It could be argued that an individual may also be curious about the topics she knows best. However, it should be noted that our definition of curiosity extends beyond the inclination of a scientist to expand her understanding within the area she is most knowledgeable about -- a characteristic that should be common to every scientist! -- but rather encompasses the search for knowledge and information far out in the knowledge space.}  

But the curiosity to explore uncharted territories creates also some tensions. In the sociology of science, this strategic tension is commonly referred to as ‘succession' versus ‘subversion' \citep{bourdieu_1975}; in organization science and innovation as ‘exploitation' versus ‘exploration' \citep{march1991exploration,gupta2006interplay}. Where do the tensions come from? Science can be viewed as a competitive territory in which scientists have to strategically choose what to study and what to cite. Compared to the returns from staying within the boundaries of the discipline, the returns from exploring other fields are systematically less certain, more distant in time, and often negative. Hence, once a scientist occupies a dominant position in a specific field, it is clear that deviating from the \textit{habitus} can be perceived as a “risky gamble”. Increasing returns from experience can trap individuals in exploiting old certainties, refining and extending existing skills, whose returns are proximate and predictable \citep{march1991exploration}. A conservative strategy allows scientists to secure publication more likely and benefit from the S\&T human capital they have accumulated. On the other hand, transcending local search space and accessing more distant knowledge opens up opportunities for originality, a prime requisite of academic reward and long-term reputation \citep{foster2015tradition}.

We posit that \textit{epistemic-specific} curiosity, or the desire for new knowledge or a particular piece of information \citep{wagstaff2021measures}, is a relevant driver of AI adoption. In the realm of AI, curiosity can arise spontaneously when some situational factors alert an individual to the existence of potential in that domain. Situational factors can be of various kind, from exposure to a sequence of events (e.g., seminars, online information) to the possession of information by someone else -- in line with our discussion in Section 2.1.1. Regardless of the specific factor, scientists with a general inclination to explore ‘new stuff' will be more likely to envision potential applications of the technology, recognize its relevance to their field of expertise, and embark on new collaborations, especially if they expect to receive high credits for the joint work.\footnote{\justifying Note that in deciding whether to collaborate, scientists generally weigh two aspects: the quality of the output and the amount of credit received for that output. Yet, the two objectives may not necessarily be aligned. Recent research in fact shows that researchers may choose to collaborate on a project even if the project is of low quality, as long as the expected benefits from the credit premium for collaboration compensate for the loss of output quality \citep{vakili2022}. In the context of AI adoption, we believe this situation could be particularly prevalent. Many domain scientists could expect recognition and credit due to the popularity of the technology, even if they are aware that the resulting scientific output may not meet high standards.}
However, as mentioned earlier, strong tensions may arise when scientists hold dominant positions within their fields, which may lead them to resist solutions and collaborations that diverge from established practices and avoid venturing beyond their comfort zone. 

Past scientific activity is a visible consequence of research choices, including a taste for exploration; and citations provide evidence of others' judgment of the relevance of a scientist's work. We will use both measures to test the following hypotheses: \\

\begin{hypothesis}
\noindent \textbf{H4a}: \textit{A higher ‘taste for exploration' increases the likelihood of adopting AI.} 
\end{hypothesis}

\begin{hypothesis}
\noindent \textbf{H4b}: \textit{A higher scientific reputation and recognition decrease the likelihood of adopting AI.} 
\end{hypothesis}

\vspace{1em}

\section{Data, variables and methods}

\subsection{Data}

Our main interest is in the adoption of AI as a research tool. More specifically, we investigate whether and how the STHC endowment of a domain scientist is related to her decision to adopt AI methods in her research. We measure adoption on scientific publications, namely i) publishing one first paper involving AI, and ii) reusing AI in at least a second paper.  

\paragraph{Source.} We use the entire OpenAlex database, which had more than 230 million scientific articles as of August 2022. OpenAlex provides us with information relevant to our study, namely titles and abstracts. Authors are disambiguated so that we can track the paper trail of our focal scientists and their co-authors. Thus, we can measure the evolution of the network of co-authors and the (publication) experience of the scientists who are part of it. Author affiliations are also clean and geographically localized. We add additional information about the organizations in our sample, particularly the university's Shanghai ranking and the availability of high-performance computing (more on this below). Finally, we rely on a scientific journal categorization system provided by OpenAlex, called ‘concept'. In particular, we denote the scientific field of a paper by 0-level concept assigned to the journal of the paper (0-level concepts include broad areas such medicine, physics, biology, chemistry, etc.)
%\footnote{\justifying There are 19 concepts at level zero, and five layers of concepts derived from them, representing a total of around 65,000 concepts. Level zero concepts include Medicine, Physics, Biology, Chemistry, etc.} 

\paragraph{Sample.} We focus on the trajectory of scientists who have used some AI in their scientific domain (other than computer science) and limit the analysis to scientists with at least two publications before the year of their first AI paper. This allows for measuring the STHC endowment of a scientist before AI adoption. In order to judge on the persistence of AI use, we also require at least one publication record after the year of the first AI-related paper. Thus, a focal scientist is observed over a period of at least three years. 

\begin{figure}[htbp!]
\centering
\includegraphics[width=0.8\textwidth]{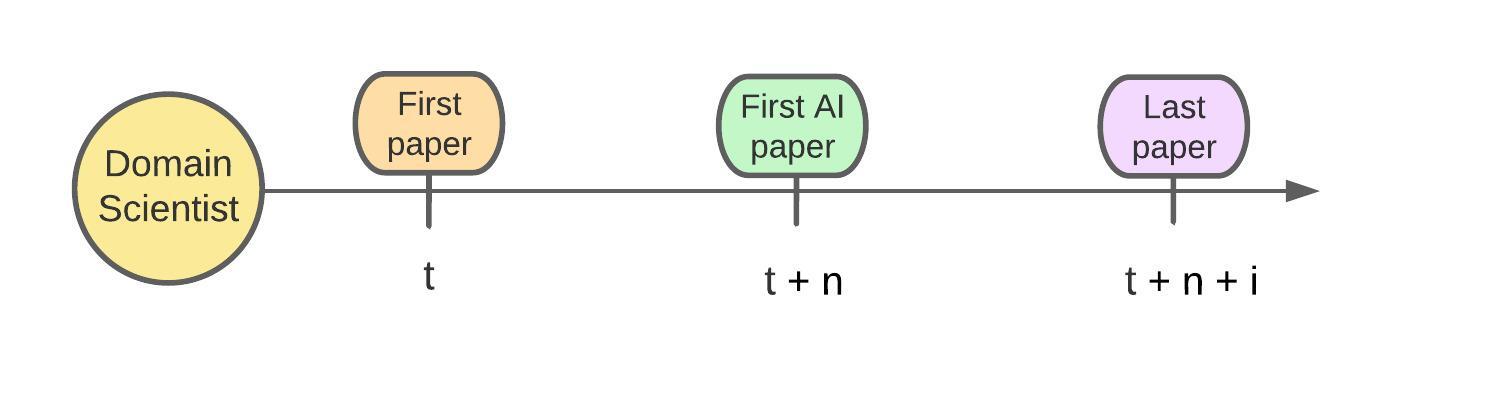}
\caption{The focal scientist. A focal scientist is active in a domain other than computer science (`domain scientist'), has a first paper in year $t>1980$, a first AI-related paper in the period $t+n=(2012,2020)$, and a subsequent last paper in the period $t+n+i=(2013,2020)$, with $n, i \geq 1$.}
\label{figure:selection}
\end{figure}

The development of AI as a research tool is a relatively recent occurrence, and the widespread adoption of deep learning applications gained momentum in the early 2010s \citep{bianchini2022artificial}. More specifically, 2012 can be considered the beginning of the AI revolution, with significant advances in neural network technology \citep{krizhevsky2017imagenet, sevilla2022compute}. This prompts us to restrict attention to the first AI use in the period from 2012 (the year when AI took off) to 2020 (end of sample period). We further limit the analysis to researchers who started research after 1980; excluding older scientists at the end of their careers and, hence, in general less susceptible to adopt AI in research. %\footnote{\justifying Researchers who began before 1980 are approaching the end of their careers, and thus their AI adoption dynamics may be less influenced by their scientific and technical human capital, and more affected by impending retirement.}
Our sampling definition of focal scientists is concisely described in Figure~\ref{figure:selection}. 

The sampling proceeds as follows. In the first step, we scanned the abstracts and titles of all papers to identify AI papers in application domains. To qualify a scientific article as an `AI paper', we build on a list of keywords provided by \cite{baruffaldi2020}. After cleaning the keyword list to preserve only terms related to AI techniques we obtain a list of 47 terms (provided in Appendix A). If any of the selected terms are mentioned in an abstract or a title of an article, then that article is considered as using AI. This approach results in a total of 1.60 million papers written by 2.83 million authors. Based on the authors' publication history, we define a scientist to be a non-computer scientist if she has no publication in a computer science journal (as indicated by OpenAlex' main concept assigned to each journal). In total, we identify 1,280,857 non-computer scientists with their first AI-related publication in the period 2012 to 2020. 

We limit our analysis to individuals who were active both before and after their exposure to AI, as illustrated in Figure \ref{figure:selection}. This ensures that a transition toward AI was made and allows us to investigate whether AI became an integral part of a researcher's toolbox (by reusing AI subsequently). To approximate STHC in the year of AI exposure, researchers must have published a minimum of two papers before that year. After conducting calculations for our main variables, as outlined in Section 3.2, and excluding authors with incomplete information, our sample reduces to 76,344 authors. These focal scientists have collectively authored 2,695,096 articles, out of which 56,733 were publications involving AI. This is the sample for our empirical analysis.%\footnote{We do not have access to individual degrees. To determine if a researcher is a computer scientist, we examine whether they have at least one publication in a journal with Computer Science as the main concept in OpenAlex. This method ensures that we identify computer scientists even if they published in non-computer science journals during the initial years of their career.}
%While this criterion may be restrictive, we believe that the ability to publish in a pure computer science journal indicates a researcher possesses specific computer science expertise.

%Our econometric strategy entails a matching approach to compare focal scientists (AI adopters) and non-focal but similar scientists that did not adopt AI. However, the implemented matching procedure makes use of various measures (similar in what?) and needs to be understood in light of the econometric strategy (why matching?). This is where we now turn. Basic descriptives of the expanded, matched sample follow at the end of this section. 

\subsection{Measures}

The following section provides details on how we measure AI adoption (the response variables), STHC endowment (explanatory variables), and further measures (control variables) for the empirical analysis.

\paragraph{AI adoption.} It is measured on scientific papers written by a focal scientist. Whether or not a paper uses AI is determined through AI keywords found in the title or abstract. Conceptually, we think of AI adoption as a process that consists of (at least) two steps. The first step is to use AI methods in research for the first time (henceforth `first-use of AI'). Then, given that first AI experience, a scientist may or may not employ AI methods subsequently (`re-use of AI'). 
\\
\paragraph{Scientific \& Technical Human Capital (STHC) endowment.} It consists of three dimensions: i) institutional capital, ii) social capital from peers, and iii) individual human capital (Figure~\ref{figure:stcap}).\\

Institutional capital is assessed by measuring a university's prestige and scientific excellence through Shanghai Ranking and citations. Institutional focus on AI is captured through AI-related papers. Availability of relevant physical infrastructure through the presence of high-performance computing (HPC) facilities. In detail, the following measures are calculated:

\begin{figure}[t!]
\includegraphics[width=\textwidth]{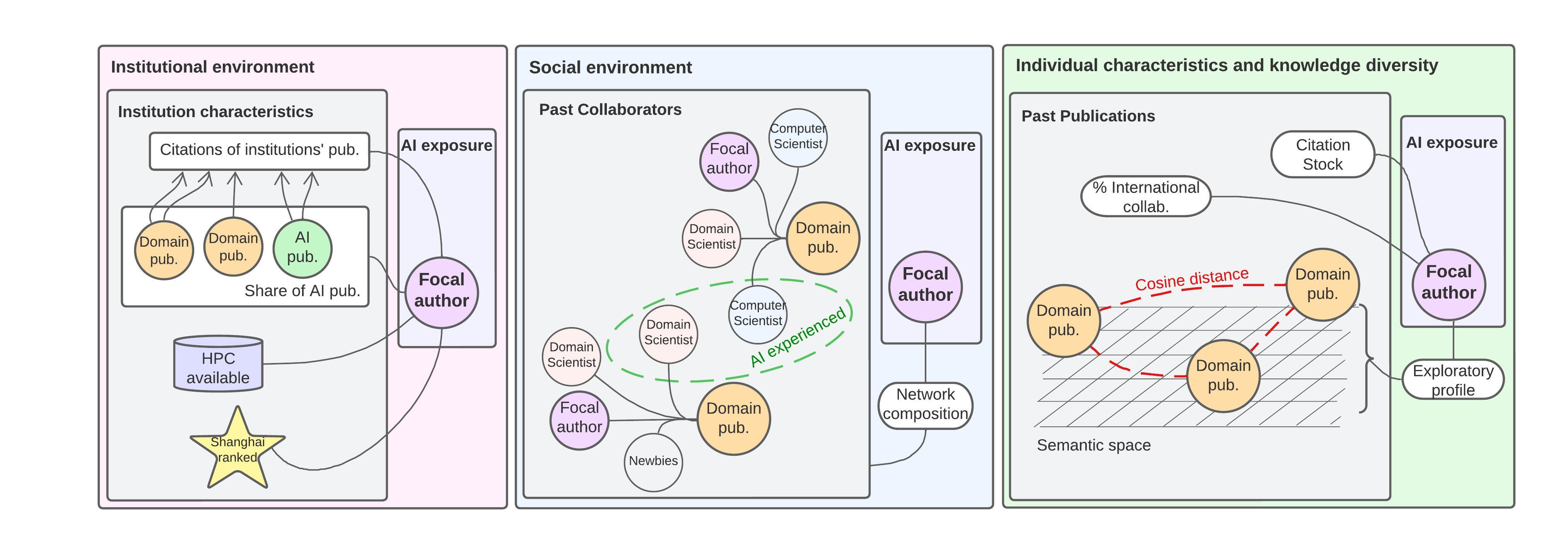}
\caption{STHC framework. \textbf{Left figure:} The institutional environment potentially provides information, directs attention, and offers resources (computing facilities, human capital) related to AI; also, institutions possess a certain level of reputation and scientific excellence. \textbf{Middle figure:} The prior co-author network provides human capital that is relevant to the focal scientists' domain, computational analysis, and/or AI. \textbf{Right figure:} The focal scientist's human capital is described by her past research output in terms of scientific content, quality, and internationality. The variables are described in detail in Section 3.2. }
\label{figure:stcap}
\end{figure}

\begin{itemize}\itemsep-0.2em
    
\item \textit{\underline{Scientific excellence}}: Scientific excellence of an institution is captured by the average number of citations per paper and year. In detail, we consider all papers in OpenAlex with at least one affiliation with that institution; for each paper, we calculate the average number of citations per year, and then average over all papers of the institution to obtain our measure of scientific excellence (\textit{`Inst. cit.'}). 

\item \textit{\underline{Prestige}}: The university's prestige is based on the Shanghai ranking. The variable \textit{`Shanghai ranked'} is used to indicate whether a specific institution is listed in the Shanghai ranking for a particular year.%\footnote{\justifying \justifying Note that our measures of scientific excellence and prestige are not imperative; there are alternative measures. Neither do they exclusively proxy these concepts, as they are also correlated with further relevant features of institutions. A university's prestige for example helps to (and allows for) the accumulation of various resources -- as more recognition brings more resources and vice versa.}  

\item \textit{\underline{AI specialization}}: If many members contribute to research using AI, it is expected that the environment is relatively supportive for conducting this type of research. Hence the degree of specialization in AI is assessed by the proportion of papers that are related to AI in a given institution and a given year. From this, we construct the binary variable \textit{`AI inst. spe.'} which takes the value one if the institution is among the top 10\% institutions in terms of degree of specialization, zero otherwise. 

\item \textit{\underline{High-Performance Computing (HPC) infrastructure}}: Access to computing resources, particularly high-performance computing (HPC) infrastructure, is often critical when conducting AI research. In order to determine whether an organization hosts an HPC infrastructure, we searched the web through the platform \textit{Perplexity} (\href{https://www.perplexity.ai}{https://www.perplexity.ai}). The platform is a sophisticated question-answering system that uses large language models to provide sourced answers to complex queries. We asked the following question: \textit{`Is there a High-Performance Computing infrastructure in the university of ... ?'}. As the answers given to our question are formulated in a very recurrent way, it was relatively easy to encode the text of the answer into `Yes' and `No' with regular expressions (we used 16 regular expressions, see Appendix A). This way, we submitted queries to Perplexity asking for HPC availability in a total of 12,500 cities and obtain an explicit indication of HPC availability ('Yes'/'No'). We manually checked 450 cities for which the platform did not provide clear indications, and found that the majority of these cities were associated with small to mid-sized universities that lacked HPC infrastructure. 

\end{itemize}

Science is a social institution in which advances depend on interactions with other scientists. Social capital in relation to peers is measured through research collaborations, which could be defined as ``\textit{[t]he working together of researchers to achieve the common goal of producing new scientific knowledge}'' (\cite{katz1997} -- p.7). Accordingly, we consider the co-author network of the focal scientist, classifying collaborators into domain scientists and computer scientists. The overall co-author network consists of 25,348,325 authors with joint papers published between 1980 and 2020. The prior co-author network in year $t$ builds on all papers (i.e., their authors and revealed co-authorship ties) from 1980 up to year $t-1$. Each scientist in the prior network is classified as a domain scientist or computer scientist based on her individual publication history. Additional co-author features taken into account are `AI experienced' and `newbie' (of the past). The social (network) capital of a focal scientist is then simply the number of prior co-authors of different types. Formally:

\begin{itemize}\itemsep-0.2em

\item \textit{\underline{Domain collaborators}}: The number of prior co-authors without any paper in a computer science journal and without any AI-related paper up to year $t-1$ (\textit{`\# Domain col.'}). 

\item \textit{\underline{Computer science collaborators}}: The number of prior co-authors with at least one paper in a computer science journal up to year $t-1$ (\textit{`\# CS col.'}). 

\item \textit{\underline{AI experienced collaborators}}: The number of prior co-authors with at least one AI-related paper up to year $t-1$ (\textit{`\# AI col.'}). 

\item \textit{\underline{Collaboration with newbies}}: The number of collaborators who had never published before the year of the collaboration (\textit{`\# Newbies col.'}). 
\end{itemize}

Note that the four types of collaborators are not mutually exclusive. For instance, an AI experienced collaborator can be either a computer scientist or a domain scientist.

Finally, individual human capital is assessed on the focal scientist's prior publications. Measures include the type of research conducted (\textit{`Scientific domain'}), the propensity to gravitate towards diverse scientific topics (\textit{`Exploratory profile'}) and the proximity to AI research (\textit{`Proximity to AI'}), recognition from past publications (\textit{`Citation stock'}), and the propensity to engage in international research collaboration (\textit{`International col.'}). 

\begin{itemize}\itemsep-0.2em

\item \textit{\underline{Scientific domain}}: The scientific domain of a domain scientist is proxied by the highest level OpenAlex concept of her first paper.

\item \textit{\underline{Exploratory profile}}: To capture an individual's ability to work on diverse topics, we compute the diversity of her papers preceding year $t$. For each paper, we represent its abstract in a vector space through word embedding methods -- i.e., Word2Vec algorithm by \cite{mikolov2013efficient}). Once these articles are positioned in the vector space, we average their pairwise cosine distance. This average distance indicates the explorative profile of the researcher.

\item \underline{\textit{Proximity to AI}}: A researcher is considered close to AI to the extent that her prior work involves OpenAlex' concepts that are frequently associated with AI. We created combinations of 284 concepts of level-1 across the entire database to asses commonness scores for each concept combination and each year, based on \cite{lee2015creativity}. The higher the commonness score, the closer the concept to AI. We list concepts of prior works of an author (up to the year of AI exposure) and select the concept most closely related to the concept `Artificial Intelligence' in the year of AI exposure. Its commonness score is our variable of interest. 

\item \textit{\underline{Citations received}}: It consists of the number of citations a focal scientist's papers received up to a year $t-1$. In detail, we use the citation count from 2022 to estimate the citation count of a given article published in year $t-x$ up to a certain year $t-1$ simply by assuming that the paper received in each year the same number of citations.

\item \textit{\underline{International collaboration}}: We measure the share of prior articles that have multiple country affiliations, which indicates the propensity of scientists to work with international peers.

\end{itemize}

\subsection{Econometric strategy}

We model the adoption of AI in the production of research papers as a combination of AI technology and STHC. An important aspect of AI research technology is that it is not a monolithic, single technology. Rather, it should be considered a bundle of various technologies undergoing specific developments, and pertaining to science specialties with differing degrees. Therefore, we allow the state of AI technology ($A$) to vary not only over time $t$ but also over the science specialty of the focal individual, denoted $s(i)$, and write $A_{s(i),t}$.

In order to usefully apply AI in research, a focal scientist $i$ may build on certain aspects of its STHC endowment accumulated up to time $t-1$, $\mathbf{H}_{i,t-1}$. Note that $\mathbf{H}$ is a vector incorporating organizational capital, social (network) capital, and individual human capital of the focal scientist. 

A scientist does not necessarily employ all STHC she is endowed with -- i.e., $\mathbf{H}_{i,t-1}$ -- to publish an individual paper. Therefore, we also consider the \textit{realized STHC} employed in a given paper $p$, denoted $\mathbf{H}_{p(i),t}$. Realized STHC ($\mathbf{H}_{p(i),t}$) will be to a large extent a part of the past STHC endowment ($\mathbf{H}_{i,t-1}$), but some capital may be acquired during research in year $t$, and some may be lost over time.

In order to fix ideas, we postulate a simple AI paper production function $F(\cdot)$ that emphasizes the complementarity (or interaction) between AI technology and the various aspects of capital:
\vspace{-0.5cm}
\begin{align*}
    F(A_{s(i),t},\mathbf{H}_{p(i),t}) = A_{s(i),t}^\gamma \; \mathbf{H}_{p(i),t}^\mathbf{\beta} 
   % &= A_{s(i),t}^\gamma \; n_{i,t}^{\beta_1} \; \exp{(\beta_2 \delta_{i,t})} \; \exp{(\nu_{i,t})}
\end{align*}

\vspace{-0.5cm}

Our main argument is that in the presence of AI certain aspects of STHC will be more valuable compared to research not dealing with AI. One immediate implication is that the realized STHC employed for an AI paper is likely to differ from that realized for a non-AI paper. Descriptive statistics at the beginning of the next section suggest that this is indeed the case.

The probability that an AI paper is produced, rather than a non-AI paper, will thus depend on the availability of those different factors of production in combination with the specific AI technology applied. For the estimation we rely on the log-transformation and a logit regression:
\vspace{-0.5cm}
\begin{align*}
    p(y_{i,t}=1) &= f(A_{s(i),t},\mathbf{H}_{i,t-1}) \\
    p(y_{i,t}=1) &= \phi \left( \gamma \log(A)_{s(i),t} + \beta \log(\mathbf{H}_{i,t-1}) + \nu_{i,t} \right) \\
    &=  \phi \left( \gamma_{s,t} + \beta \mathbf{h}_{i,t-1} + \nu_{i,t} \right)
\end{align*}
\vspace{-1cm}

\noindent where $p(y_{i,t}=1)$ is the probability that a paper produced applies AI conditional on a paper produced at all, $\phi$ denotes the logit-function. $\mathbf{h}_{i,t-1}$ is the log-transform of our measurements $\mathbf{H}_{i,t-1}$. In principle, some relevant capital and/or individual specific tendencies may be also unobserved $\nu_{i,t}$.%\footnote{\justifying For convenience, the coefficients associated with AI research technology and STHC have the same names in the paper production function and the adoption function, but they are of course not the same.}
With this estimation equation, the dynamics of AI technologies is effectively dealt with by introducing intercepts $\gamma_{s,t}$ that vary over time and scientific domains. 

Our estimation strategy is based on a matching approach where we match scientists belonging to the same scientific field and same cohort, but with different adoption behaviour: one group integrates AI, the other does not. Conditional logit regressions on the matched pairs removes all their common factors from the estimation regression. One common factor is the AI technology itself, another are cohort specific (unobserved) human capital and preferences. 
As outlined above, it seems likely that scientists in the same field and belonging to the same cohort face similar (exogenous) dynamic of AI technology $A_{s(i),t}$. Furthermore, scientists of same field and cohort may share some similarities in unobserved preferences and skills, $\nu_{s(i),t}$. This leads us to write individual unobserved components as the sum of average cohort effects $\bar{\nu}_{s(i),t}$ and individual deviations from that average $\tilde{\nu}_{i,t}$ -- i.e., $\nu_{i,t} = \bar{\nu}_{s(i),t} + \tilde{\nu}_{i,t}$. By matching same cohort scientists $i$ and $j$, we obtain $A_{s(i),t} = A_{s(j),t}$ and $ \bar{\nu}_{s(i),t} =  \bar{\nu}_{s(j),t}$.

These common factors can be removed in a conditional logit approach. First note that matching is on the outcome such that in all matches one individual $i$ adopts AI and the other scientist $j$ does not adopt AI $(y_{i,t}=1, y_{j,t}=0)$, or vice versa. The conditional logit model, takes into account that only two possible outcomes are possible, and we estimate the probability of one of them,  where common factors have been pulled out of the second equation:

\vspace{-0.5cm}
\begin{align*}
p(y_{i,t}=1,y_{j,t}=0) &= \frac{exp( \gamma_{s,t} + \beta h_{i,t-1} + \bar{\nu}_{s(i),t} + \tilde{\nu}_{i,t} )}{exp( \gamma_{s,t} + \beta h_{i,t-1} + \bar{\nu}_{s(i),t} + \tilde{\nu}_{i,t} )+exp( \gamma_{s,t} + \beta h_{j,t-1} + \bar{\nu}_{s(j),t} + \tilde{\nu}_{j,t} )} \\
	&= \frac{exp( \beta h_{i,t-1} +  \tilde{\nu}_{i,t} )}{exp(  \beta h_{i,t-1} + \tilde{\nu}_{i,t} )+exp(  \beta h_{j,t-1} + \tilde{\nu}_{j,t} )} 
\end{align*}
\vspace{-0.5cm}

Our regressions control for the scientific background of the focal scientist in different ways, such as through scientific domain fixed effects and other variables that measure her exploratory profile and proximity to AI. Yet, in case further individual but unobserved components $\tilde{\nu}_{i,t}$ are correlated with observed factors in $\mathbf{h}$, coefficient estimates can be biased. Consider for example an individual's (hidden) preference for data-intensive research, influencing collaboration with computer scientists and shaping a specific social capital type. If this unobserved aspect is not factored in the model, the impact of this social capital on AI adoption probability would be exaggerated. 

With this caveat in mind, our strategy is not to precisely identify the marginal impact of different STHC factors on technology adoption. Adoption is neither exogenous nor random in our data; it is a choice. We observe correlations between STHC on one hand and AI adoption on the other, and we interpret these patterns in light of the conceptual framework developed in Section 2.

% NO NEEDED IN VIEWPOINT -- 
%This must be taken into account in the interpretation of the estimation results; we are discussing correlations, not causation. 
%Note that in the estimation equation above the variation of AI technology is dealt with as a simple scaling factor of the valuation of STHC. A more flexible and probably more realistic formulation would be to allow for variation of the exponents ($\beta$'s) of STHC with the dynamics of AI technologies. For example, computer scientists may be indeed needed in the early stages of AI development, whereas in, say, two generations domain scientists may well be capable of autonomously using AI (because of both different human capital formed during training and different AI technology). We investigate the variation of STHC coefficients across scientific fields and time by estimating additional regressions.

\subsubsection{Matching}

Technology available at a time $t$, age of the researcher, initial training and research trajectory influence the opportunity to use AI. Hence, when matching individuals with similar potential to use AI, we take into account technology advancement and its applicability in a given field.

\begin{figure}[t!]
\centering
\includegraphics[width=0.7\textwidth]{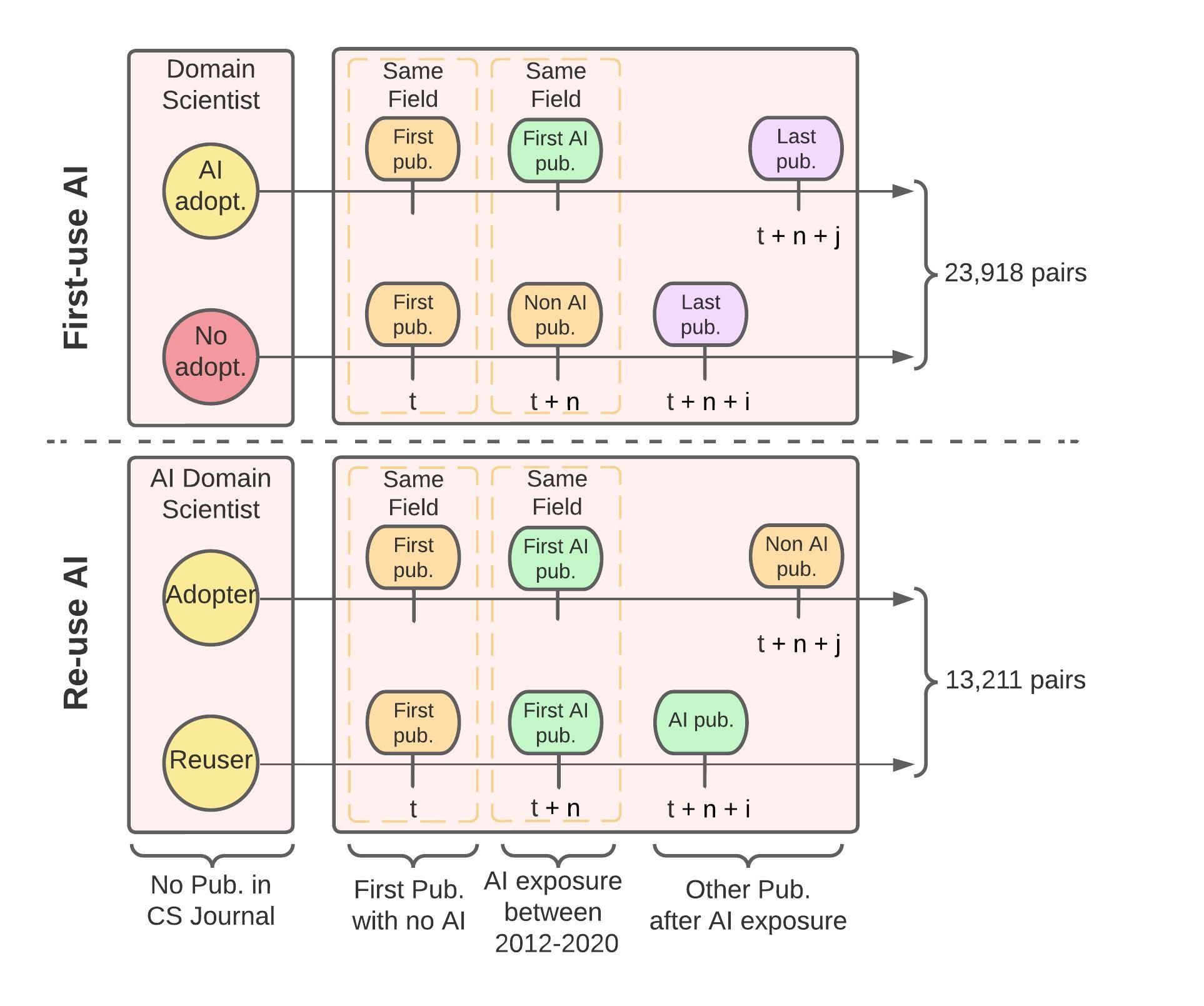}
\caption{Matching procedures to investigate first-use of AI (top) and re-use of AI (bottom)}
\label{figure:matching}
\end{figure}

As shown in Figure \ref{figure:matching}, we construct two matched samples. The first matching is used to investigate first-use of AI. Each focal scientist is matched with a non-focal scientist who (i) never published an AI paper, but published her first paper in the same year and same scientific field as the focal scientist, (ii) published a non-AI paper in the same scientific field and same year as the focal scientists' first AI paper, and (iii) published at least one paper subsequently, as the focal scientist. Since our focus is on domain scientists, individuals with computer science papers are excluded (focal and matched scientist). 

The second matching is used to investigate re-use of AI. Here we compare two focal scientists who first published a paper on AI in the same scientific field and in the same year, but one of them reuses AI in a later year and the other does not. As for the first matching, we also require that both individuals published their first paper in the same field and year.

We performed exact matching using all Openalex data, matching 23,918 first-users of AI with non-users of AI and connecting 13,211 re-users with first-users that did not re-use AI.

\section{Empirical analysis}

This section provides at first some aggregate statistics on the diffusion of AI in research and discusses the main patterns of its adoption process. Detailed descriptive statistics are available in Appendix A. The section then focuses on the main econometric results regarding the relationship between STHC endowment, first-use of AI and its subsequent re-use. The analysis is broken down by scientific field in Appendix B.

\subsection{Main trends on AI adoption in science}

%In a nutshell the number of scientists with first-time AI use is strongly increasing in particular since 2017, with a stable share of around one third re-using AI subsequently. There is also a strong cohort effect such that in particular early career scientists adopt AI.
%Turning to the factors affecting AI adoption, we see that i) focal scientists co-author more frequently with computer scientists and AI experienced scientists in AI-related papers than in non-AI papers (Table~\ref{tab:RealizedSTHC}), and that ii) AI adopters, be it first-time AI use or re-use of AI, tend to have a larger endowment of AI relevant STHC than those that do not adopt AI (Table~\ref{tab:STHCMatchedSamples}).
%The last two tables, Tables~\ref{tab:corrFirsttime} and \ref{tab:corrReuse}, provide correlations among all the variables on the matched samples used to estimate the effect of STHC endowment on first-time AI use and re-use of AI respectively.

\noindent \underline{\textit{A growing number of players}}: The adoption of AI technology by domain scientists is on the rise, with over 20,000 individuals incorporating it for the first time in 2020 alone. This pattern aligns with recent research on the diffusion of AI in science, showing that the number of AI publications has surged nearly five-fold in the past decade, constituting around 5\% of total scientific publications. Most of this research has gradually shifted from AI development to its application, which currently represents some 70\% of scientific activity \citep{ec2023}. Similar patterns have been observed for AI-related patent applications in OECD and non-OECD countries \citep{parteka2023}. Surprisingly, despite technological advancements and increased resources for AI, the percentage of first-time users reusing technology remains low and stable at around 54\% (13,211 re-use out of 23,918 first-use).

Our sample is comprised mostly of early- and mid-career scientists, indicating lower involvement of older generations in AI research. On average, the first year of publication is 2006, and focal scientists are exposed to AI at an average academic age of 11.05 years, with a median of 9 years. \\

\noindent \underline{\textit{AI adoption seems to require different teams}}: A basic tenet underlying our study is that AI research is realized with specific STHC that differs from STHC used in non-AI research. Ideally, one would like to observe the extent to which a focal scientist with a given STHC endowment leverages each dimension of this STHC for different research projects. This would allow to see which aspects of STHC are particularly relevant when it comes to AI. However, our publication data usually lacks such detailed information. The exception is the social capital dimension in the form of co-authors with their individual scientific background.

\begin{table}[t!] \centering 
\renewcommand{\arraystretch}{1}
\setlength{\tabcolsep}{10pt}
  \caption{Co-authors of first-time AI adopters} 
  \label{tab:RealizedSTHC} 
\begin{threeparttable}
\begin{tabular}{llll}
\toprule
 & non-AI papers & AI papers & $t$-test\\
\midrule
\# Authors & 10.73 (10.31) & 11.65 (12.51) & 14.32$^{***}$\\
\# CS aut. & 2.19 (4.09) & 3.12 (4.82) & 36.66$^{***}$\\
\# AI exp. aut. & 1.71 (3.48) & 2.84 (4.54) & 49.34$^{***}$\\
\# Domain aut. & 8.53 (7.65) & 8.53 (9.62) & 0.04\\
\# Newbies aut. & 1.19 (1.70) & 1.41 (2.82) & 16.13$^{***}$\\
\midrule
Observations & 62,712 & 62,712 & \\
\bottomrule
\end{tabular}
\begin{tablenotes}
 \footnotesize
 \item \justifying   {\it Notes:}
 This table presents descriptive statistics on realized STHC for AI adopters. It compares an AI paper with a non-AI paper published by focal scientists in their first year of AI exposure. The table provides averages and standard deviations (in parentheses) of the number of different types of co-authors. $t$-tests determine whether the mean differences between the groups are statistically significant. Significance levels are denoted by ***, **, and * for the 1\%, 5\%, and 10\% levels, respectively. Note that for this table focal scientists without a non-AI paper in the year of their first-time AI paper are excluded.
 \end{tablenotes}
 \end{threeparttable}

\end{table}

Table~\ref{tab:RealizedSTHC} compares co-authors of non-AI papers with co-authors of AI papers published by focal scientists in the year of their first-use of AI. The sample includes for each focal scientist one (randomly chosen) non-AI paper and one (randomly chosen) AI paper, both written by the focal scientist and published in the year of the focal scientist's first-time AI use. On average, focal scientists work with one more co-author in AI papers (11.65 authors in total) compared to non-AI papers (10.73 authors in total). Interestingly, this additional co-author typically has a background in computer science. Moreover, the research team is more likely to include AI experienced scientists. On the other hand, the number of domain (non-computer) scientists is on average the same. Finally, newbies without any prior publication contribute somewhat more often to AI papers than to non-AI papers. Thus, team composition varies by AI use for the same scientist (with a given STHC).\\

%The lower part of the table (`co-authors and citations of matched papers') provides additional evidence on AI specific resource employment. When engaging in AI (first-use and re-use) focal scientists tend to work in teams of a different composition than their peers not engaging in AI, i.e. less domain co-authors and more AI experienced and computer science co-authors. Their number of newbies (first-time publishers) is not significantly different however.{\color{red} TODO Are these newbies that have never published before, or tagged as newbies because they had a first paper with the focal in the past?}\\

\noindent \underline{\textit{AI adopters are endowed with more STHC}}: Table \ref{tab:STHCMatchedSamples} presents descriptive statistics for the two samples employed in the regression analysis of first-use of AI (left) and re-use (right).

\begin{table}[t!] \centering \centering \footnotesize
\renewcommand{\arraystretch}{1}
\setlength{\tabcolsep}{0.1pt}
  \caption{Descriptive statistics for matched samples} 
  \label{tab:STHCMatchedSamples}
\scalebox{0.84}{
\begin{threeparttable}
\begin{tabular}{@{\extracolsep{2pt}}l  c c c | c c c} 
\toprule
 & \multicolumn{3}{c}{First-use AI} & \multicolumn{3}{c}{Re-use AI} \\[5pt]
& Matched scientists & Focal scientists & $t$-test & Matched scientists & Focal scientists & $t$-test\\
 & (without AI) & (with AI) &  & (not re-using AI) & (re-using AI) & \\
\midrule
\underline{STHC endowment} & \multicolumn{1}{c}{} & \multicolumn{1}{c}{} & \multicolumn{1}{c}{} & \multicolumn{1}{c}{} & \multicolumn{1}{c}{} & \multicolumn{1}{c}{} \\
\addlinespace
AI inst. spec. & 0.05 (0.22) & 0.10 (0.29) & 19.07$^{***}$ & 0.08 (0.27) & 0.09 (0.29) & 2.73$^{***}$\\
Inst. cit. & 2.86 (3.88) & 3.04 (1.98) & 6.57$^{***}$ & 3.09 (2.09) & 3.14 (2.03) & 2.06$^{**}$\\
Shanghai ranked & 0.03 (0.17) & 0.03 (0.17) & -0.09 & 0.03 (0.17) & 0.03 (0.17) & -0.47\\
HPC & 0.71 (0.45) & 0.74 (0.44) & 6.16$^{***}$ & 0.73 (0.44) & 0.74 (0.43) & 2.58$^{***}$\\
\addlinespace
\# Domain col. & 146.84 (181.62) & 155.58 (220.97) & 4.73$^{***}$ & 134.27 (190.6) & 167.48 (245.87) & 12.27$^{***}$\\
\# CS col. & 21.08 (33.88) & 27.83 (44.68) & 18.6$^{***}$ & 23.86 (40.68) & 31.46 (49.54) & 13.62$^{***}$\\
\# AI col. & 9.30 (18.43) & 14.56 (25.89) & 25.62$^{***}$ & 12.36 (24.12) & 17.18 (29.74) & 14.46$^{***}$\\
\# Newbies col. & 51.95 (68.31) & 55 (81.34) & 4.45$^{***}$ & 46.83 (69.11) & 59.05 (89.77) & 12.41$^{***}$\\
\addlinespace
Exploratory profile & 0.18 (0.06) & 0.18 (0.06) & 2.29$^{**}$ & 0.18 (0.06) & 0.18 (0.06) & 3.01$^{***}$\\
Citation stock & 918.2 (2040.94) & 1031.02 (2441.89) & 5.48$^{***}$ & 826.93 (1959.27) & 1088.13 (2625.3) & 9.16$^{***}$\\
\% International pub. & 0.30 (0.25) & 0.30 (0.25) & -1.73$^{*}$ & 0.30 (0.26) & 0.29 (0.25) & -0.95\\
Proximity to AI & 0.20 (0.24) & 0.26 (0.28) & 22.78$^{***}$ & 0.59 (1.25) & 0.79 (1.48) & 11.59$^{***}$\\
\midrule
\underline{Co-authors and citations of matched papers} & \multicolumn{1}{c}{} & \multicolumn{1}{c}{} & \multicolumn{1}{c}{} & \multicolumn{1}{c}{} & \multicolumn{1}{c}{} & \multicolumn{1}{c}{} \\
\addlinespace
\# Domain aut. & 10.88 (9.99) & 8.40 (8.9) & -28.68$^{***}$ & 8.38 (8.96) & 7.67 (7.26) & -7.04$^{***}$\\
\# AI exp. aut. & 1.31 (2.68) & 2.77 (4.25) & 44.9$^{***}$ & 2.67 (4.38) & 2.87 (4.39) & 3.7$^{***}$\\
\# CS aut. & 1.78 (3.46) & 3.00 (4.45) & 33.42$^{***}$ & 2.99 (4.73) & 3.07 (4.55) & 1.36\\
\# Newbies aut. & 1.28 (2.86) & 1.24 (2.12) & -1.63 & 1.23 (2.21) & 1.21 (2.17) & -0.66\\
\# Citations & 4.15 (13.49) & 4.75 (10.82) & 5.37$^{***}$ & 4.54 (10.48) & 5.43 (12.2) & 6.37$^{***}$\\
\midrule
Total & 23,918 & 23,918 & & 13,211 & 13,211 &  \\
\bottomrule
\end{tabular} 
\begin{tablenotes}
 \item \justifying   {\it Notes:}
This table presents the descriptive statistics for various variables, including their mean values and standard deviations (in parentheses) for different matched samples. The table also provides results from $t$-tests to determine if the mean differences between the groups are statistically significant. Significance levels are denoted by ***, **, and * for the 1\%, 5\%, and 10\% levels, respectively.
 \end{tablenotes}
 \end{threeparttable}
 }
\end{table}

The upper part of Table \ref{tab:STHCMatchedSamples} presents STHC endowment of sampled scientists. The $t$-tests indicate that nearly all variables representing STHC endowment show significant mean differences between individuals who apply AI and those who abstain. At the institutional level, the percentage of individuals working in specialized institutions is two times larger (\textit{AI inst. spec.}), and the citation impact of these institutions appears to be positively correlated with AI adoption (\textit{Inst. cit.}). This is true for first-use of AI and, to a lesser extent, for re-using AI. Whether or not the university is listed in the Shanghai ranking seems to make no difference.%\footnote{ \justifying \justifying One potential explanation is that our measure indicating whether the university is among the top 1000 universities is too rough to capture existing `elite-effects'.} 

Social capital is also richer for individuals who adopt AI technology: more domain scientists, more computer scientists, more collaborators with AI experience and more early-career researchers. Finally, at the individual level, prior accomplishments (citation stock) and proximity to AI also appear to be positively linked to AI adoption.

Examining the lower part of Table~\ref{tab:STHCMatchedSamples}, AI papers authored by first-time AI users (left part of the table) exhibit a composition of co-authors that is different to papers of matched individuals who did not adopt AI. Specifically, the number of domain scientists tends to be smaller for first-time AI users, while the presence of individuals with computer science specialization and AI experience nearly doubles. Furthermore, it can be observed that individuals that have previously worked on concepts closer to AI concepts are generally more inclined to engage with AI technology and subsequently incorporate it into other articles.

We observe a similar pattern when examining the co-authors in papers of re-users and non-re-users (right part of the table). The citation count of the initial AI article is higher among those who persist in using AI technology. Moreover, individuals who incorporate AI into their subsequent research already possess a more specialized STHC endowment in computer science and AI compared to other researchers who will not continue using the technology. They are also more likely to be affiliated with highly specialized AI institutions and have increased access to computing centers. Ultimately, individuals who maintain the use of AI in their research exhibit a more exploratory profile and demonstrate stronger past success than their counterparts who do not continue. \\

% MOVED TABLES TO APPENDIX TO SLIM DOWN THE TEXT -- Tables \ref{tab:corrFirsttime} and \ref{tab:corrReuse} display basic descriptive statistics for our two estimation samples. We can immediately observe that the measures derived from collaboration networks are highly correlated among each other. In particular, the number of newbie collaborators and the number of domain scientist collaborators are strongly correlated with a Pearson correlation coefficient of above 0.9 in both estimation samples. This is however expected because many newbies are likely to count as domain scientists. Similarly, the number of collaborators with AI experience is strongly related to the number of collaborators in computer science; in both samples the correlation coefficient is above 0.8. Additionally, a clear positive correlation is visible between the number of collaborations and the stock of citations. For completeness, the last variable included in each table is the outcome variable, i.e. first-time AI use and re-use of AI respectively. Note however that Table~\ref{tab:STHCMatchedSamples} is more appropriate to shed light on the relation between our (binary) left-hand-side variable and right-hand-side variables. \\

\noindent \underline{\textit{Researchers use AI predominantly within their specialized domains}}: Researchers adopting AI tend to apply this technology in the field they know best, particularly where their initial research was published. More than 60\% of researchers publish their first AI article in the journal of the same scientific field as their first publication. The majority of researchers in our sample published in life sciences-related journals, with approximately 80\% of first-time AI papers appearing in the fields of medicine, biology, and chemistry. The next section explores more depth the adoption process in six specific fields (Medicine, Biology, Chemistry, Physics, Psychology, and Materials Science), which together constitute 95\% of our sample.

\subsection{Econometric results}

\subsubsection{First-time AI adoption}
We present here the results of the conditional logit regression of first-time AI use on three dimensions of scientific and technical human capital (STHC): institutional, social, and individual factors. Recall that in the estimation sample the outcome, first-time AI adoption, is one for focal scientists and zero for matched scientists. STHC is measured up to the year before adoption.

\begin{table}[t!]
\centering
\renewcommand{\arraystretch}{0.8}
\setlength{\tabcolsep}{0.3pt}
  \caption{Conditional Logit with matching (first-time AI adoption)} 
  \label{tab:adopting} 
\scalebox{0.87}{
\begin{threeparttable}
\begin{tabular}{@{\extracolsep{35pt}}l c c c c } 
\toprule
% & \multicolumn{4}{c}{\textit{Dependent variable: First-time AI adoption}} \\ 
%\cline{2-5} 
\\[-1.8ex] & \multicolumn{1}{c}{Institutional} & \multicolumn{1}{c}{Social} & \multicolumn{1}{c}{Individual} & \multicolumn{1}{c}{Full Model} \\ 
\\[-1.8ex] & \multicolumn{1}{c}{(1)} & \multicolumn{1}{c}{(2)} & \multicolumn{1}{c}{(3)} & \multicolumn{1}{c}{(4)}\\ 
\hline \\[-1.8ex] 
AI inst. spe. & 0.723$^{***}$ &  &  & 0.591$^{***}$ \\ 
 & (0.039) &  &  & (0.043) \\ 
 & & & & \\ 
Inst. cit. & 0.413$^{***}$ &  &  & 0.248$^{***}$ \\ 
 & (0.029) &  &  & (0.033) \\ 
 & & & & \\ 
Shanghai ranked & -0.009 &  &  & 0.011 \\ 
 & (0.058) &  &  & (0.065) \\ 
 & & & & \\ 
HPC & 0.065$^{***}$ &  &  & 0.001 \\ 
 & (0.021) &  &  & (0.024) \\ 
 & & & & \\ 
\# Domain col. &  & -1.503$^{***}$ &  & -1.376$^{***}$ \\ 
 &  & (0.034) &  & (0.036) \\ 
 & & & & \\ 
\# CS col.  &  & 0.139$^{***}$ &  & 0.127$^{***}$ \\ 
 &  & (0.020) &  & (0.021) \\ 
 & & & & \\ 
\# AI col. &  & 0.793$^{***}$ &  & 0.774$^{***}$ \\ 
 &  & (0.019) &  & (0.019) \\ 
 & & & & \\ 
\# Newbies col. &  & 0.575$^{***}$ &  & 0.511$^{***}$ \\ 
 &  & (0.025) &  & (0.026) \\ 
 & & & & \\ 
Exploratory profile &  &  & 0.053 & 1.574$^{***}$ \\ 
 &  &  & (0.168) & (0.193) \\ 
 & & & & \\ 
Citation stock &  &  & -0.073$^{***}$ & -0.082$^{***}$ \\ 
 &  &  & (0.007) & (0.011) \\ 
 & & & & \\ 
\% International pub. &  &  & 0.006 & -0.140$^{***}$ \\ 
 &  &  & (0.039) & (0.046) \\ 
 & & & & \\ 
Proximity to AI &  &  & 1.393$^{***}$ & 0.840$^{***}$ \\ 
 &  &  & (0.056) & (0.064) \\ 
 & & & & \\ 
\hline \\[-1.8ex] 
Observations & \multicolumn{1}{c}{47,836} & \multicolumn{1}{c}{47,836} & \multicolumn{1}{c}{47,836} & \multicolumn{1}{c}{47,836} \\ 
%R$^{2}$ & \multicolumn{1}{c}{0.013} & \multicolumn{1}{c}{0.096} & \multicolumn{1}{c}{0.015} & \multicolumn{1}{c}{0.107} \\ 
%Max. Possible R$^{2}$ & \multicolumn{1}{c}{0.500} & \multicolumn{1}{c}{0.500} & \multicolumn{1}{c}{0.500} & \multicolumn{1}{c}{0.500} \\ 
Log Likelihood & \multicolumn{1}{c}{-16,259.650} & \multicolumn{1}{c}{-14,171.280} & \multicolumn{1}{c}{-16,229.090} & \multicolumn{1}{c}{-13,882.300} \\ 
%Wald Test & \multicolumn{1}{c}{595.610$^{***}$ (df = 4)} & \multicolumn{1}{c}{3,583.870$^{***}$ (df = 4)} & \multicolumn{1}{c}{661.950$^{***}$ (df = 4)} & \multicolumn{1}{c}{3,891.570$^{***}$ (df = 12)} \\ 
LR Test & \multicolumn{1}{c}{638.085$^{***}$ } & \multicolumn{1}{c}{4,814.822$^{***}$} & \multicolumn{1}{c}{699.209$^{***}$} & \multicolumn{1}{c}{5,392.798$^{***}$ } \\ 
%Score (Logrank) Test & \multicolumn{1}{c}{623.236$^{***}$ (df = 4)} & \multicolumn{1}{c}{4,314.416$^{***}$ (df = 4)} & \multicolumn{1}{c}{686.349$^{***}$ (df = 4)} & \multicolumn{1}{c}{4,781.234$^{***}$ (df = 12)} \\ 
\bottomrule
\end{tabular} 
\begin{tablenotes}
 \footnotesize
 \item \justifying   {\it Notes:} This table reports coefficients of the effect STHC on first-time AI adoption on all fields. ***, ** and * indicate significance at the 1\%, 5\% and 10\% level, respectively. The effect of STHC on AI adoption is estimated using a conditional logit with matching. \end{tablenotes}
 \end{threeparttable}
 }
\end{table} 

Table \ref{tab:adopting} provides coefficient estimates of four models --- one column for each dimension of STHC separately and the fourth column for estimating all coefficients jointly. The log-likelihood ratio tests (LR  Test) at the bottom of the table confirm that all models improve significantly over the intercept-only model. Considering the log-likelihood in increasing order, we see that social factors (Column 2) are the most informative dimension for first-time AI adoption, consistent with the idea that AI-relevant STHC does not solely reside in the individual scientist but in particular in her social endowments.

\bigskip

\noindent \underline{\textit{Institutional factors}}:
The positive and significant coefficients of \textit{AI inst. spe.} and \textit{Inst. cit.} in models (1) and (4) suggest that scientists affiliated with institutions specialized in AI research and with higher citation impact are more likely to adopt AI in their work, validating H1b. Being affiliated with a Shanghai-ranked institution does not significantly correlate with first-use of AI. 

%When researchers are part of a highly specialized AI institution, it is a matter of time before they use AI in their research. 
As put forward in Section 2, researchers are heavily influenced by their institutional environment, including shared norms, directives, and funding focused on specific subjects. As such, an organization with a strong AI focus and recognized research output may offer fertile ground to facilitate AI adoption. 

Taken at face value the estimated coefficients suggest that being affiliated with an institution specialized in AI (\textit{AI inst. spe.}=1), rather than not (\textit{AI inst. spe.}=0), increases the odds-ratio for the focal scientist adopting AI compared to the matched scientist by 80\%.\footnote{\justifying To appreciate this point, note that our conditional logit model estimated on matched pairs allows for exactly two outcomes: either the focal scientist adopts AI and the matched scientist does not, or vice versa. The odds-ratio of the focal scientist adopting AI over the matched scientist adopting AI is $exp(\beta h_i)/exp(\beta h_j)$. A change of $i$'s factor from $h_i$ to $h^{\prime}_{i}$ thus implies a relative change of the odds-ratio by $exp(\beta (h^{\prime}_{i} - h_i))$. A binary factor changing from 0 to 1 yields a relative change of $exp(\beta)$, and a continuous factor (in logs in the model) being higher by 10\% is about $exp(\beta \times 1.1)$. Hence, interpretation of coefficient estimates is easily facilitated by considering the exponential of the coefficient estimate.}
Scientists of more ‘excellent' institutions with 10\% higher citation rates have 30\% higher odds-ratios of AI adoption. 

We reject H3 since access to high-performance computing (HPC) resources is significantly associated with AI adoption only in model (1) but not in the full model (4). This suggests that HPC may not be crucial, overall. One reason could be that many state-of-the-art AI models are available pre-trained and hence using them for inferencing does not necessarily require elevated computational resources (which may be instead relevant for training such models). Further analysis of the effect of HPC by scientific domain -- estimates are provided in Appendix A -- reveals that computing resources can be an important driver of adoption and re-use only in certain domains, particularly chemistry and medicine.\\

\noindent \underline{\textit{Social factors}}:
Model (2) and full model (4) show that the number of domain collaborators (\textit{\# Domain col.}) has a negative and significant effect on AI adoption, suggesting that a more extensive network of domain collaborators may reduce, as expected, the likelihood to adopt. On the other hand, the number of computer science collaborators (\textit{\# CS col.}) and collaborators with previous AI experience (\textit{\# AI col.}) both have positive and significant effects, thereby confirming H1a. 

Consistent with the above discussion, it seems that prior contacts with computer scientists and/or AI-experienced scientists help acquire about-knowledge on AI, thus enabling the researcher to more easily integrate teams using AI. At the same time, researchers with too many domain-specific collaborators may be less exposed to `AI-related thinking' or be part of niches where AI is still not widely used. The positive influence of early-career collaborators (\textit{\# Newbies col.}) on adoption confirms H2. Individuals can indeed expand their expertise through the fresh perspectives of young scientists, who are more likely to be better trained in modern statistical and computational methods. 

Although one must be cautious in interpreting these estimates as causal effects (recall Section 3.3), they certainly delineate an upper range of social (over-)embeddedness in AI adoption that is substantial: 10\% more domain collaborators reduce the odds-ratio of AI adoption by nearly 80\%. On the other hand, 10\% more collaborators of other types are associated with substantially higher odds-ratios of AI adoption -- more computer scientists by 15\%; more AI experienced collaborators by 130\%; and early-career collaborators (newbies) by 80\%. These estimates, obtained on the whole sample, are broadly consistent across scientific disciplines (see Appendix B).\\

\noindent \underline{\textit{Individual Factors}}: As shown in Models (3) and (4), we find empirical support to H4a: a higher taste for exploration positively influences AI adoption. This might be due to the fact that individuals with a more diversified cognitive profile are more accustomed to exploring subjects that are far from their prior knowledge and, in the case of AI, possibly reorienting their research to seize the opportunities offered by the technology. Individuals who have engaged in concepts often associated with AI (\textit{Proximity to AI}) are more inclined to adopt the technology. Citation stock is negatively correlated with adoption, confirming H4b and suggesting that individuals with a higher scientific reputation and recognition are actually less likely to integrate AI into their research. Lastly, international co-authorship (\textit{\% International pub.}) reduces the likelihood of first-time AI adoption. This could imply that researchers who tend to work more with local colleagues have stronger connections that facilitate better knowledge transfer, in particular when the environment is rich of AI-relevant resources, thereby contributing more effectively to developing a reasonable understanding of AI.

In terms of odds-ratios of AI adoption, an exploratory profile with a 10\% higher score or 10\% more proximity to AI are associated with about 10\% higher odds-ratios. 10\% more citations or international papers reduce odds-ratios by 10 and 15\% respectively.

Note, however, that there is some heterogeneity across scientific fields (see Appendix B) and, at the current stage of our research, we do not have sufficiently strong empirical arguments to claim an explanation for the specific effects of individual factors.

\subsubsection{AI re-use}

Table \ref{tab:reusing} shows the effect of realized STHC employed in the first AI paper (upper part) and STHC endowment (lower part) on the re-use of AI in subsequent research. The four models mirror those used in the regression of first-time AI use, but now the sample is limited to focal scientists with an AI publication. Once more, log-likelihood tests confirm significant improvements over the intercept-only model, with social factors continuing to offer the best fit (Column 2), followed by individual (Column 3) and institutional factors (Column 1). The full model (Column 4) again controls for all available factors.

\begin{table}[h!] \centering 
\renewcommand{\arraystretch}{0.85}
\setlength{\tabcolsep}{0.1pt}
  \caption{Conditional logit with matching (re-using AI)} 
  \label{tab:reusing} 
\scalebox{0.86}{
\begin{threeparttable}
\begin{tabular}{@{\extracolsep{40pt}}lc c c c} 
\toprule
% & \multicolumn{4}{c}{\textit{Dependent variable: re-using AI}} \\ 
%\cline{2-5} 
\\[-1.8ex] & \multicolumn{1}{c}{Institutional} & \multicolumn{1}{c}{Social} & \multicolumn{1}{c}{Individual} & \multicolumn{1}{c}{Full Model} \\ 
\\[-1.8ex] & \multicolumn{1}{c}{(1)} & \multicolumn{1}{c}{(2)} & \multicolumn{1}{c}{(3)} & \multicolumn{1}{c}{(4)}\\ 
\hline \\[-1.8ex] 
\\
\underline{\textit{Realized STHC}} \\
\\
\hspace{0.5em} \# Domain aut. & -0.303$^{***}$ & -0.330$^{***}$ & -0.295$^{***}$ & -0.298$^{***}$ \\ 
  & (0.026) & (0.028) & (0.026) & (0.028) \\ 
  & & & & \\ 
\hspace{0.5em} \# AI exp. aut. & 0.173$^{***}$ & 0.063$^{**}$ & 0.172$^{***}$ & 0.061$^{**}$ \\ 
  & (0.024) & (0.027) & (0.024) & (0.027) \\ 
  & & & & \\ 
\hspace{0.5em} \# CS aut. & -0.055$^{**}$ & -0.113$^{***}$ & -0.073$^{***}$ & -0.107$^{***}$ \\ 
  & (0.024) & (0.026) & (0.024) & (0.026) \\ 
  & & & & \\ 
\hspace{0.5em} \# Newbies aut.  & 0.155$^{***}$ & 0.184$^{***}$ & 0.177$^{***}$ & 0.183$^{***}$ \\ 
  & (0.024) & (0.025) & (0.024) & (0.025) \\ 
  & & & & \\ 
\hspace{0.5em} \# Citations& 0.151$^{***}$ & 0.165$^{***}$ & 0.136$^{***}$ & 0.166$^{***}$ \\ 
  & (0.016) & (0.016) & (0.016) & (0.016) \\ 
  & & & & \\ 
\\
\underline{\textit{STHC endowment}} \\
\\  
\hspace{0.5em} AI inst. spe.& 0.102$^{**}$ &  &  & 0.112$^{**}$ \\ 
  & (0.046) &  &  & (0.047) \\ 
  & & & & \\ 
\hspace{0.5em} Inst. cit.& 0.114$^{***}$ &  &  & -0.016 \\ 
  & (0.039) &  &  & (0.041) \\ 
  & & & & \\ 
\hspace{0.5em} Shanghai ranked  & -0.016 &  &  & -0.051 \\ 
  & (0.077) &  &  & (0.079) \\ 
  & & & & \\ 
\hspace{0.5em} HPC & 0.051$^{*}$ &  &  & 0.038 \\ 
  & (0.029) &  &  & (0.030) \\ 
  & & & & \\ 
\hspace{0.5em} \# Domain col. &  & -0.258$^{***}$ &  & -0.223$^{***}$ \\ 
  &  & (0.038) &  & (0.040) \\ 
  & & & & \\ 
\hspace{0.5em} \# CS col.  &  & 0.079$^{***}$ &  & 0.075$^{***}$ \\ 
  &  & (0.027) &  & (0.028) \\ 
  & & & & \\ 
\hspace{0.5em} \# AI col.  &  & 0.274$^{***}$ &  & 0.272$^{***}$ \\ 
  &  & (0.025) &  & (0.025) \\ 
  & & & & \\ 
\hspace{0.5em} \# Newbies col. &  & 0.245$^{***}$ &  & 0.210$^{***}$ \\ 
  &  & (0.030) &  & (0.030) \\ 
  & & & & \\ 
\hspace{0.5em} Exploratory profile &  &  & 0.970$^{***}$ & 0.796$^{***}$ \\ 
  &  &  & (0.233) & (0.242) \\ 
  & & & & \\ 
\hspace{0.5em} Citation stock  &  &  & 0.118$^{***}$ & 0.004 \\ 
  &  &  & (0.010) & (0.014) \\ 
  & & & & \\ 
\hspace{0.5em} \% International pub. &  &  & -0.138$^{***}$ & -0.322$^{***}$ \\ 
  &  &  & (0.052) & (0.056) \\ 
  & & & & \\ 
\hspace{0.5em} Proximity to AI &  &  & 0.330$^{***}$ & 0.222$^{***}$ \\ 
  &  &  & (0.028) & (0.029) \\ 
  & & & & \\ 
\hline \\[-1.8ex] 
Observations & \multicolumn{1}{c}{26,422} & \multicolumn{1}{c}{26,422} & \multicolumn{1}{c}{26,422} & \multicolumn{1}{c}{26,422} \\ 
%R$^{2}$ & \multicolumn{1}{c}{0.011} & \multicolumn{1}{c}{0.033} & \multicolumn{1}{c}{0.023} & \multicolumn{1}{c}{0.037} \\ 
%Max. Possible R$^{2}$ & \multicolumn{1}{c}{0.500} & \multicolumn{1}{c}{0.500} & \multicolumn{1}{c}{0.500} & \multicolumn{1}{c}{0.500} \\ 
Log Likelihood & \multicolumn{1}{c}{-9,007.784} & \multicolumn{1}{c}{-8,717.901} & \multicolumn{1}{c}{-8,854.444} & \multicolumn{1}{c}{-8,658.276} \\ 
%Wald Test & \multicolumn{1}{c}{288.380$^{***}$ (df = 9)} & \multicolumn{1}{c}{797.650$^{***}$ (df = 9)} & \multicolumn{1}{c}{563.950$^{***}$ (df = 9)} & \multicolumn{1}{c}{894.120$^{***}$ (df = 17)} \\ 
LR Test & \multicolumn{1}{c}{298.766$^{***}$ } & \multicolumn{1}{c}{878.533$^{***}$ } & \multicolumn{1}{c}{605.448$^{***}$} & \multicolumn{1}{c}{997.784$^{***}$ } \\ 
%Score (Logrank) Test & \multicolumn{1}{c}{295.204$^{***}$ (df = 9)} & \multicolumn{1}{c}{849.853$^{***}$ (df = 9)} & \multicolumn{1}{c}{590.914$^{***}$ (df = 9)} & \multicolumn{1}{c}{960.713$^{***}$ (df = 17)} \\ 
\bottomrule
\end{tabular} 
\begin{tablenotes}
 \footnotesize
 \item \justifying   {\it Notes:} This table reports coefficients of the effect of STHC endowment and realized STHC on re-using AI. ***, ** and * indicate significance at the 1\%, 5\% and 10\% level, respectively. The effect of STHC on re-using AI is estimated using a conditional logit with matching. \end{tablenotes}
 \end{threeparttable}
 }
\end{table}

The estimates for the STHC endowment closely resemble those for the first-use of AI across all dimensions in terms of direction and statistical significance, but tend to be significantly smaller in magnitude. The citation stock of researchers and their institutions no longer appear influential in the pursuit of AI research. Taken together, this suggests that a scientist's STHC endowment continues to play its role also in re-using AI, but is less discriminating among scientists having experienced first-time AI use. 

We now turn to how characteristics of the first AI paper relate to re-using AI (realized STHC in the upper part of Table 4). First note the positive influence through the rewards linked to the first AI publication, particularly in terms of citations (\textit{\# Citations}). Further variables capture how the team composition of the first AI paper relates to re-using AI in subsequent research: A higher prevalence of individuals with previous AI experience within a team (\textit{\# AI exp. aut.}) denotes the team's specialization in AI. And it is, indeed, within this specialized social context that researchers enhance their likelihood of reusing AI in subsequent work. Across all models (1-4), we observe that the number of domain experts (\textit{\# Domain aut.}) negatively affects AI reuse, while the presence of novice researchers (\textit{\# Newbies aut.}) plays a positive role. The negative role of domain authors, and the positive role of AI experienced authors and newbies in the first AI paper (in realized STHC) are all in line with the estimates obtained for the different co-author types in the prior collaboration network (STHC endowment). 

The results on the role of computer scientists are somewhat peculiar. Regarding the first AI paper, the presence of computer scientists in the prior collaboration network (\textit{CS col.}) is positively associated with AI reuse. But the presence of computer scientists as co-authors in the first AI paper (\textit{CS aut.}) plays exactly the opposite effect. %\footnote{\justifying In order to ensure that this result does not arise from ignoring non-linear effects of individual variables or their interaction, we tried various alternative specifications with and without squared terms and with varying sets of variables. This result appeared to be robust, and even stronger when adding squared terms.} 
A domain scientist collaborating with, say, three computer scientists in the first AI paper, compared to no computer scientists in the team, has an approximately 10\% lower odds-ratio for AI reuse. Although the estimated association is of relatively small magnitude compared to other factors, a brief discussion seems warranted. We propose the following explanation. Collaboration with computer scientists is likely to be instrumental in the diffusion of AI, as multi-disciplinary teams seem well suited to apply and adapt AI tools from computer sciences to research problems in other domains. Our data, indeed, support the relevance of collaboration between computer scientists and researchers of other domains in AI adoption. It is also clear, however, that such interdisciplinary teams may face various hurdles in their formation and subsequent life-cycle \citep[see, e.g.,][]{Fiore2008}. For instance, in their analysis of scientific collaboration among Stanford faculty members, \cite{DahlanderMcfarland2013} find that intellectual dissimilarity has a negative effect not only on first-tie formation but also on collaboration persistence. Similarly, our findings show that collaboration between computer scientists and domain scientists that led to the first AI paper is in general difficult to maintain and is unlikely to lead to persistent AI adoption by domain scientists.

%At the same time, the very observation of such an interdisciplinary collaboration potentially signals its necessity. Our observations are consistent with the idea that a high dependence on computer scientists can be resolved in individual research projects through interdisciplinary collaboration, but --- due to the fragility of interdisciplinary collaboration --- may not result in persistent AI adoption by domain scientists. 

\section{Discussion and conclusion}

In this study, we examined different factors influencing the adoption and reuse of AI in scientific research, focusing on institutional, social, and individual factors. We sought to provide empirical evidence on what are the main drivers in the adoption of new technologies for scientific discovery, a problem hitherto not addressed in the literature.

First, we highlighted the importance of AI specialization within institutions: When AI ‘is in the air', the likelihood of adoption increases, consistent with our conceptual framework. Second, and contrary to our expectations, access to high-performance computing (HPC) resources and affiliation with top-tier universities may not be decisive factors, at least in most fields of application (except, for example, medicine and chemistry). Third, regarding social factors, we showed that network position matters, particularly when scientists are closely connected to peers who have already used AI in their past research. Also, collaborations with early-career researchers -- i.e., newbies -- seem to contribute significantly to the adoption process. Finally, we found that some individual characteristics are important when it comes to integrating AI into scientific practices. A taste for exploration, for instance, seems to enhance the ability of individuals to recognize the potential of AI in their application domains and prompt them to ‘give it a try'; conversely, having a dominant position and high reputation within a field tends to hinder this propensity. Taken together, our findings offer some insights for policymakers and science administrators aiming to enhance the diffusion of AI tools in the sciences, providing them with a broader understanding of the complex interplay between these factors. Some critical reflections are therefore in order.

Let us begin by reconsidering the institutional framework within which research is conducted. An organizational climate that emphasizes individual competition over cooperation may pose a barrier to knowledge sharing and circulation. Indeed, given the relative ease with which funding for AI research can be obtained (at least at present), scientists may be reticent to share their AI-related knowledge with their colleagues to avoid intensifying competition. Incentives can be put in place to create a supportive culture and foster knowledge circulation within institutions as well as among epistemic communities. Such incentives can be intrinsic, such as recognition and praise, and further supplemented with extrinsic rewards, such as bonuses and higher salaries. 

While incentives are valuable, facilitating knowledge sharing can also be achieved by establishing a work environment that promotes interactions among scientists and communication across departments. This can be accomplished, for instance, through informal AI-focused events rather than (often futile) interdisciplinary ambitions. In this regard, organizations should be able to identify ‘boundary-spanning’ individuals who are eager to share their (tacit) AI knowledge and expertise with their peers, while also possessing effective communication skills to engage non-AI experts and pique their interest. Lastly, we think that organizations could set some research priorities around AI, thus creating a sense of group identity and personal responsibility \citep{cabrera2002knowledge}. A relevant theory for this purpose is that of ‘organization-based self-esteem’ (OBSE), which refers to the degree to which an individual considers him/herself capable, significant, and worthy as a member of the organization \citep{wang2010knowledge}. As such, scientists may be more likely to share their knowledge with others if they feel their competencies align with the organization’s goals.

Enthusiasm for interdisciplinary collaborations is very visible in the policy agenda of recent years. Although initiatives aimed at bringing together the AI community and scientists from specific application areas can be helpful, we should not forget that collaboration is first and foremost an individual choice, a decision that involves a number of important trade-offs and comes with some costs -- e.g., coordination and credit allocation \citep{bikard2015}. We believe that the costs associated with (interdisciplinary) collaborations could be especially high in the context of AI because, as shown here, the integration of AI tools often requires teams with diverse cultural backgrounds, each having their unique incentives, research priorities, and cultural norms. Policy should encourage not only interdisciplinary teamwork, but also ensure that this collaboration is sustained in the medium-long term; this is critical to create an environment that embraces the possibility of failure, which often occurs when researchers from different fields collaborate for the first time.

Our results seem to suggest that computational resources are not a major determinant of AI adoption, except for a few areas. Thus, if the policy ambition is to democratize AI in as many application domains as possible, one may wonder whether large investments in computational facilities such as HPC – which come with substantial overheads and the need for specialized human resources – are the most effective strategy. Alternatively, a more modest but widespread investment in data science/ML laptops and workstations can be a powerful vehicle for AI adoption in the sciences and, why not, a mechanism to broaden access to technology and close computing divides.

We do not rule out, however, that computational resources are a significant asset when it comes to cutting-edge AI research, as evidence suggests \citep{sevilla2022compute}. According to a recent study by the \cite{/content/paper/876367e3-en}, when asked about the main barriers or challenges to accessing AI computation, about 50\% of respondents cited the cost of AI compute. Thus, the lack of financial resources for most public and private organizations can give a group of big players -- i.e., elite universities and large tech companies -- an unfair advantage that results in a concentration of power. More in general, we believe that further research is needed to understand better the demand for AI compute, particularly in domain-specific applications, and not solely for core AI research.

Finally, we refrain from making recommendations on how to nudge individual choices toward exploration while maintaining some degree of exploitation, which is nonetheless essential for scientific progress. Yet we can state with some confidence that current trends in science policy and scientific communities, from impact assessments to targeted research funding (see, e.g., \cite{franzoni2022funding}), are unlikely to favor exploratory research paths. Perhaps it is time to rethink these habits as well.

%If this is indeed the case, interdisciplinary collaboration with computer scientists should be considered a transitory device in AI adoption; to be employed with the aim to integrate AI skills into the set of basic research capabilities underlying any application domain. 

\clearpage

\bibliography{refs}
\newpage

%%%%% APPENDIX %%%%%%

\clearpage

\appendix
\section*{Appendix}
\renewcommand{\thesubsection}{\Alph{subsection}}

\subsection{Data and methods}

\setcounter{table}{0}
\renewcommand{\thetable}{A\arabic{table}}

\begin{table}[!h] \centering \centering 
\renewcommand{\arraystretch}{1}
\setlength{\tabcolsep}{0.5pt}
  \caption{Number of authors per OpenAlex's concepts} 
  \label{} 
\scalebox{0.84}{
\begin{threeparttable}
\begin{tabular}{@{\extracolsep{65pt}}l r r} 
\toprule
Concept & First AI publ. & First publ.\\
\midrule
Medicine & 29,762 & 31,823\\
Biology & 29,469 & 23,200\\
Physics & 5,727 & 5,995\\
Chemistry & 5,691 & 7,525\\
Psychology & 2,549 & 2,489\\
\addlinespace
Materials science & 1,763 & 1,856\\
Geology & 1,213 & 1,085\\
Economics & 1,080 & 414\\
Engineering & 856 & 905\\
Geography & 357 & 297\\
\addlinespace
Mathematics & 180 & 243\\
Political science & 103 & 187\\
Philosophy & 54 & 114\\
Environmental science & 41 & 53\\
Business & 32 & 49\\
\addlinespace
Sociology & 17 & 34\\
Art & 10 & 57\\
History & 3 & 18\\
\midrule
Total & 78,907 & 76,344\\
\bottomrule
\end{tabular}
\begin{tablenotes}
 \footnotesize
 \item \justifying    {\it Notes:} This table reports the number of authors per OpenAlex's concept for their first AI publication and their first publication in the sample.
 \end{tablenotes}
 \end{threeparttable}
 }
\end{table}

\begin{table} \centering
\caption{List of terms used to flag AI papers in OpenAlex}
\renewcommand{\arraystretch}{.9}
\scalebox{0.93}{
\begin{threeparttable}
\begin{tabular}{l r}
\toprule
Term & Count [\%] \\
\midrule
 neural network &  719,132 [44.93] \\
                  machine learning &   345,707 [21.60] \\
                     deep learning &  180,207 [11.26] \\
           artificial intelligence &   156,690 [9.79] \\
         artificial neural network &   147,165 [9.19] \\
                    vector machine &   143,682 [8.98] \\
            support vector machine &   140,205 [8.76] \\
      convolutional neural network &   115,962 [7.24] \\
               deep neural network &    64,263 [4.01] \\
                     random forest &    63,662 [3.98] \\
            reinforcement learning &    61,592 [3.85] \\
               supervised learning &    57,715 [3.61] \\
                           k-means &    53,397 [3.34] \\
       natural language processing &    47,590 [2.97] \\
                         markovian &    36,139 [2.26] \\
          recurrent neural network &    35,604 [2.22] \\
                  bayesian network &    29,888 [1.87] \\
                 transfer learning &     24,012 [1.50] \\
                   backpropagation &     23,938 [1.50] \\
               adversarial network &    20,700 [1.29] \\
                       autoencoder &    20,442 [1.28] \\
    generative adversarial network &     19,229 [1.20] \\
             unsupervised learning &    18,666 [1.17] \\
 deep convolutional neural network &    18,229 [1.14] \\
         support vector regression &    14,639 [0.91] \\
                   regression tree &    14,048 [0.88] \\
               stochastic gradient &    11,346 [0.71] \\
            multi-layer perceptron &      9,665 [0.60] \\
                        q-learning &      9,577 [0.60] \\
          semi-supervised learning &      9,558 [0.60] \\
                 ensemble learning &     9,091 [0.57] \\
       latent dirichlet allocation &      6,416 [0.40] \\
                 bayesian learning &      4,765 [0.30] \\
            long short term memory &     4,212 [0.26] \\
    natural language understanding &     4,195 [0.26] \\
             variational inference &     3,828 [0.24] \\
          latent semantic analysis &     3,616 [0.23] \\
               deep belief network &     3,581 [0.22] \\
                   kernel learning &     3,091 [0.19] \\
       natural language generation &     2,551 [0.16] \\
                  hebbian learning &     2,416 [0.15] \\
           instance-based learning &      718 [0.04] \\
       nearest neighbour algorithm &      563 [0.04] \\
                     neural turing &      101 [0.01] \\
             neural turing machine &       99 [0.01] \\
          extreme machine learning &       86 [0.01] \\
\bottomrule
\end{tabular}
%\begin{tablenotes}
% \footnotesize
% \item \justifying   {\it Notes:} This table reports the list of AI terms used to identify AI articles.
% \end{tablenotes}
 \end{threeparttable}
 }
\end{table}

\begin{table}
\centering 
\renewcommand{\arraystretch}{1}
\setlength{\tabcolsep}{0.5pt}
\caption{Regular expressions used to label HPC availability}
\scalebox{0.84}{
\begin{threeparttable}
\begin{tabular}{@{\extracolsep{7pt}}l l} 
\toprule
Label & Regular Expression\\
\midrule
Yes & \\
& `\textasciicircum yes, ' \\
& `\textasciicircum the .\{0,50\} has a high( \textbar -)performance computing' \\
& `\textasciicircum the .\{0,50\} does have a high( \textbar -)performance computing' \\
& `\textasciicircum the university of .\{0,50\} has an infrastructure for high( \textbar -)performance computing' \\
& `\textasciicircum there is a high( \textbar -)performance computing' \\
& `\textasciicircum the university of .\{0,50\} has a computational infrastructure' \\
& `it also has a high( \textbar -)performance computing' \\
\midrule
No & \\
& `\textasciicircum no, ' \\
& `\textasciicircum there is no information available' \\
& `\textasciicircum the university of .\{0,50\} is not mentioned' \\
& `there is no evidence' \\
& `there is no mention of' \\
& `there is no evidence that the university .\{0,50\} has a high( \textbar -)performance computing infrastructure' \\
& `it is unclear (whether\textbar if) the' \\
& `\textasciicircum it is not clear if the university of .\{0,50\} has a high( \textbar -)performance computing' \\
& `does not appear' \\
& `\textasciicircum the university of .\{0,50\} does not have a high' \\
& `\textasciicircum there is no clear information' \\
& `i could not find .\{0,4\} information'\\
& 'does not have its high performance'\\
& `it appears that the university of .\{0,50\} does not have'\\
& `\textasciicircum there is no information in the provided search'\\
\bottomrule
\end{tabular}
\begin{tablenotes}
 \footnotesize
 \item \justifying   {\it Notes:} This table reports the regular expressions used to label HPC availability in a given institution based on the answer received from Perplexity after asking the query: `Is there a High-Performance Computing infrastructure in the University of ...? '
 \end{tablenotes}
 \end{threeparttable}
 }
\end{table}

\clearpage
\newpage

\begin{table}[!ht] \centering
\renewcommand{\arraystretch}{0.3}
\setlength{\tabcolsep}{4pt}
  \caption{Descriptive statistics and correlation matrix (sample for first-time AI adoption)}  
  \label{tab:corrFirsttime} 
\scalebox{0.8}{
\begin{threeparttable}
\begin{tabular}{lrrllllllllllll}
\toprule
Variable & Mean & Std &  (1) &(2) &(3) &(4) &(5) &(6) &(7) &(8) &(9) &(10) &(11) &(12)\\
\midrule
(1) AI inst. spe. & 0.07 & 0.26 & 1 &  &  &  &  &  &  &  &  &  &  &  \\
(2) Inst. cit. & 1.31 & 0.35 & 0.03 & 1 &  &  &  &  &  &  &  &  &  &  \\
(3) Shanghai ranked & 0.03 & 0.17 & 0.01 & 0.02 & 1 &  &  &  &  &  &  &  &  &  \\
(4) HPC & 0.72 & 0.45 & 0.01 & 0.22 & -0.01 & 1 &  &  &  &  &  &  &  &   \\
(5) \# Domain col. & 4.41 & 1.14 & -0.11 & 0.03 & -0.03 & 0.04 & 1 &  &  &  &  &  &  &  \\
(6) \# CS col. & 2.55 & 1.19 & -0.04 & 0.16 & -0.03 & 0.11 & 0.79 & 1 &  &  &  &  &   &  \\
(7) \# AI col. & 1.83 & 1.17 & -0.04 & 0.17 & -0.05 & 0.11 & 0.69 & 0.84 & 1 &  &   &  &  &  \\
(8) \# Newbies col. & 3.33 & 1.21 & -0.08 & -0.02 & -0.01 & 0.00 & 0.93 & 0.68 & 0.56 & 1 &  &  &  &   \\
(9) Exploratory profile & 0.18 & 0.06 & -0.09 & -0.1 & -0.01 & -0.02 & 0.26 & 0.13 & 0.11 & 0.26 & 1 &  &  &  \\
(10) Citation stock & 5.54 & 1.80 & -0.03 & 0.15 & -0.01 & 0.09 & 0.75 & 0.70 & 0.56 & 0.70 & 0.07 & 1  &  &  \\
(11) \% International pub. & 0.30 & 0.25 & 0.01 & 0.09 & -0.02 & 0.01 & 0.18 & 0.27 & 0.26 & 0.09 & -0.16 & 0.13 & 1 &    \\
(12) Proximity to AI & 0.19 & 0.19 & -0.02 & 0.01 & -0.02 & 0.04 & 0.31 & 0.39 & 0.34 & 0.31 & 0.12 & 0.26 & 0.00 & 1 \\
(13) First-time AI (yes/no) & 0.50 & 0.50 & 0.09 & 0.07 & 0.00 & 0.03 & -0.06 & 0.08 & 0.15 & -0.04 & 0.01 & -0.02 & -0.01 & 0.1 \\
\bottomrule
\end{tabular}
 \end{threeparttable}
 }
\end{table}

\vspace{5em}

\begin{table}[!h] \centering \centering 
\renewcommand{\arraystretch}{1.2}
\setlength{\tabcolsep}{3.4pt}
  \caption{Descriptive statistics and correlation matrix (sample for AI re-use)} 
  \label{tab:corrReuse} 
\scalebox{0.66}{
\begin{threeparttable}
\begin{tabular}{lrrlllllllllllllllll}
\toprule
Variable & Mean & Std &  (1) &(2) &(3) &(4) &(5) &(6) &(7) &(8) &(9) &(10) &(11) &(12) &(13) &(14) &(15) &(16) &(17)\\
\midrule
(1) AI inst. spe. & 0.09 & 0.28 & 1 &  &  &  &  &  &  &  &  &  &   &  &  &  &  &  & \\
(2) Inst. cit.  & 1.35 & 0.34 & 0.00 & 1 &  &  &  &  &  &  &  &  &   &  &  &  &  &  & \\
(3) Shanghai ranked  & 0.03 & 0.17 & -0.01 & 0.00 & 1 &  &  &  &  &    &  &  &  &  &  &  &  &  & \\
(4) HPC  & 0.74 & 0.44 & -0.03 & 0.21 & -0.03 & 1 &  &  &  &  &  &  &  &  &  &  &  &  & \\
(5) \# Domain col. & 4.25 & 1.30 & -0.14 & 0.06 & -0.02 & 0.04 & 1 &  &  &  &  &  &  &  &  &  &  &  &   \\
(6) \# CS col. & 2.61 & 1.25 & -0.08 & 0.17 & -0.03 & 0.1 & 0.81 & 1 &  &  &  &  &  &  &  &  &  &  &   \\
(7) \# AI col.  & 2.00 & 1.21 & -0.09 & 0.17 & -0.05 & 0.1 & 0.72 & 0.86 & 1 &  &  &  &  &  &  &  &  &  &   \\
(8) \# Newbies col. & 3.18 & 1.34 & -0.11 & 0.01 & -0.01 & 0.01 & 0.94 & 0.71 & 0.6 & 1 &  &  &  &  &  &  &  &  &   \\
(9) Exploratory profile & 0.18 & 0.06 & -0.11 & -0.05 & 0.00 & 0.00 & 0.34 & 0.2 & 0.17 & 0.33 & 1 &  &  &  &  &  &  &  &   \\
(10) Citation stock  & 5.35 & 1.92 & -0.07 & 0.15 & -0.02 & 0.08 & 0.79 & 0.74 & 0.61 & 0.74 & 0.16 & 1 &  &  &  &  &  &  &  \\
(11) \% International pub. & 0.30 & 0.25 & 0.01 & 0.07 & -0.02 & 0.00 & 0.16 & 0.26 & 0.25 & 0.09 & -0.12 & 0.12 & 1 &  &  &  &  &  &  \\
(12) Proximity to AI & 0.36 & 0.48 & -0.01 & 0.03 & -0.02 & 0.04 & 0.23 & 0.32 & 0.28 & 0.24 & 0.04 & 0.24 & 0.03 & 1 &  &  &  &  &  \\
(13) \# Domain aut. & 1.98 & 0.62 & -0.08 & 0.02 & -0.02 & -0.01 & 0.31 & 0.17 & 0.22 & 0.26 & 0.14 & 0.12 & 0.13 & -0.06 & 1 &  &  &  &  \\
(14) \# AI exp. aut. & 0.99 & 0.77 & -0.02 & 0.11 & -0.03 & 0.07 & 0.14 & 0.27 & 0.41 & 0.06 & 0.02 & 0.08 & 0.16 & 0.06 & 0.32 & 1 &  &  &  \\
(15) \# CS aut. & 1.08 & 0.75 & -0.01 & 0.14 & -0.03 & 0.08 & 0.14 & 0.35 & 0.33 & 0.08 & 0.03 & 0.10 & 0.19 & 0.09 & 0.22 & 0.70 & 1 &  & \\
(16) \# Newbies aut.& 0.57 & 0.61 & 0.01 & 0.00 & 0.01 & -0.05 & 0.06 & -0.04 & -0.01 & 0.09 & 0.05 & -0.03 & 0.03 & -0.06 & 0.48 & 0.04 & 0.07 & 1 & \\
(17) \# Citations & 1.33 & 0.85 & 0.03 & 0.17 & 0.00 & 0.02 & 0.00 & 0.07 & 0.05 & -0.01 & -0.04 & 0.09 & 0.07 & 0.01 & 0.09 & 0.19 & 0.23 & 0.01 & 1 \\
(18) Re-use AI (yes/no) & 0.50 & 0.50 & 0.02 & 0.03 & 0.00 & 0.02 & 0.05 & 0.10 & 0.12 & 0.06 & 0.02 & 0.06 & -0.01 & 0.08 & -0.04 & 0.04 & 0.03 & 0.01 & 0.06 \\
\bottomrule
\end{tabular}
 \end{threeparttable}
 }
\end{table}

\newpage
\setcounter{table}{0}
\renewcommand{\thetable}{B\arabic{table}}

\subsection{Extension: Field-level analysis}

\begin{table}[h!] \centering
\renewcommand{\arraystretch}{0.8}
\setlength{\tabcolsep}{0.3pt}
  \caption{Conditional logit with matching across fields (first-time AI adoption)} 
  \label{tab:adoptingfield} 
\scalebox{0.84}{
\begin{threeparttable}
\begin{tabular}{@{\extracolsep{15pt}}l c c c c c c } 
\toprule
% & \multicolumn{6}{c}{\textit{Dependent variable: first-time AI use}} \\ 
% \cline{2-7} 
\\[-1.8ex] & \multicolumn{1}{c}{Medicine} & \multicolumn{1}{c}{Biology} & \multicolumn{1}{c}{Chemistry} & \multicolumn{1}{c}{Physics} & \multicolumn{1}{c}{Psychology} & \multicolumn{1}{c}{Materials science} \\ 
\\[-1.8ex] & \multicolumn{1}{c}{(1)} & \multicolumn{1}{c}{(2)} & \multicolumn{1}{c}{(3)} & \multicolumn{1}{c}{(4)} & \multicolumn{1}{c}{(5)} & \multicolumn{1}{c}{(6)}\\ 
\hline \\[-1.8ex] 
 AI inst. spe. & 0.663$^{***}$ & 0.644$^{***}$ & 0.296$^{**}$ & 0.588$^{***}$ & 0.716$^{**}$ & 0.784$^{***}$ \\ 
  & (0.083) & (0.069) & (0.125) & (0.144) & (0.298) & (0.185) \\ 
  & & & & & & \\ 
 Inst. cit. & 0.322$^{***}$ & 0.265$^{***}$ & -0.216 & 0.261$^{*}$ & 0.177 & 0.854$^{***}$ \\ 
  & (0.051) & (0.053) & (0.133) & (0.143) & (0.241) & (0.293) \\ 
  & & & & & & \\ 
  Shanghai ranked & 0.038 & 0.043 & -0.037 & -0.499 & -0.330 & -0.090 \\ 
  & (0.110) & (0.099) & (0.222) & (0.360) & (0.384) & (0.460) \\ 
  & & & & & & \\ 
 HPC & -0.029 & 0.004 & 0.268$^{***}$ & 0.108 & 0.083 & -0.169 \\ 
  & (0.038) & (0.038) & (0.087) & (0.104) & (0.177) & (0.184) \\ 
  & & & & & & \\ 
 \# Domain col. & -1.214$^{***}$ & -1.418$^{***}$ & -1.663$^{***}$ & -1.610$^{***}$ & -1.385$^{***}$ & -1.269$^{***}$ \\ 
  & (0.059) & (0.057) & (0.141) & (0.145) & (0.220) & (0.266) \\ 
  & & & & & & \\ 
 \# CS col. & 0.085$^{**}$ & 0.212$^{***}$ & 0.102 & 0.033 & -0.188 & 0.513$^{***}$ \\ 
  & (0.034) & (0.034) & (0.075) & (0.094) & (0.156) & (0.153) \\ 
  & & & & & & \\ 
 \# AI col. & 0.793$^{***}$ & 0.659$^{***}$ & 0.678$^{***}$ & 1.097$^{***}$ & 1.278$^{***}$ & 0.572$^{***}$ \\ 
  & (0.031) & (0.030) & (0.071) & (0.083) & (0.133) & (0.130) \\ 
  & & & & & & \\ 
 \# Newbies col. & 0.537$^{***}$ & 0.536$^{***}$ & 0.517$^{***}$ & 0.326$^{***}$ & 0.290$^{*}$ & 0.120 \\ 
  & (0.042) & (0.041) & (0.095) & (0.103) & (0.167) & (0.176) \\ 
  & & & & & & \\ 
 Exploratory profile & 2.048$^{***}$ & 1.104$^{***}$ & 1.798$^{**}$ & 2.666$^{**}$ & 2.279$^{**}$ & -1.482 \\ 
  & (0.289) & (0.309) & (0.869) & (1.131) & (1.112) & (1.938) \\ 
  & & & & & & \\ 
Citation stock & -0.087$^{***}$ & -0.096$^{***}$ & -0.020 & -0.048 & -0.105 & -0.203$^{**}$ \\ 
  & (0.017) & (0.018) & (0.043) & (0.041) & (0.074) & (0.084) \\ 
  & & & & & & \\ 
 \% International pub.  & -0.249$^{***}$ & -0.025 & -0.094 & -0.436$^{**}$ & 0.590$^{**}$ & 0.044 \\ 
  & (0.074) & (0.073) & (0.175) & (0.206) & (0.296) & (0.340) \\ 
  & & & & & & \\ 
 Proximity to AI  & 1.143$^{***}$ & 0.604$^{***}$ & 1.185$^{***}$ & 0.380$^{*}$ & 0.254 & 0.362 \\ 
  & (0.104) & (0.103) & (0.262) & (0.231) & (0.332) & (0.481) \\ 
  & & & & & & \\ 
\hline \\[-1.8ex] 
Observations & \multicolumn{1}{c}{19,684} & \multicolumn{1}{c}{18,484} & \multicolumn{1}{c}{3,400} & \multicolumn{1}{c}{2,758} & \multicolumn{1}{c}{1,374} & \multicolumn{1}{c}{1,022} \\ 
R$^{2}$ & \multicolumn{1}{c}{0.103} & \multicolumn{1}{c}{0.101} & \multicolumn{1}{c}{0.123} & \multicolumn{1}{c}{0.178} & \multicolumn{1}{c}{0.188} & \multicolumn{1}{c}{0.144} \\ 
%Max. Possible R$^{2}$ & \multicolumn{1}{c}{0.500} & \multicolumn{1}{c}{0.500} & \multicolumn{1}{c}{0.500} & \multicolumn{1}{c}{0.500} & \multicolumn{1}{c}{0.500} & \multicolumn{1}{c}{0.500} \\ 
Log Likelihood & \multicolumn{1}{c}{-5,756.889} & \multicolumn{1}{c}{-5,422.077} & \multicolumn{1}{c}{-955.048} & \multicolumn{1}{c}{-686.307} & \multicolumn{1}{c}{-333.513} & \multicolumn{1}{c}{-274.914} \\ 
%Wald Test (df = 12) & \multicolumn{1}{c}{1,567.480$^{***}$} & \multicolumn{1}{c}{1,443.260$^{***}$} & \multicolumn{1}{c}{303.730$^{***}$} & \multicolumn{1}{c}{308.500$^{***}$} & \multicolumn{1}{c}{161.470$^{***}$} & \multicolumn{1}{c}{103.810$^{***}$} \\ 
LR Test  & \multicolumn{1}{c}{2,130.132$^{***}$} & \multicolumn{1}{c}{1,967.979$^{***}$} & \multicolumn{1}{c}{446.603$^{***}$} & \multicolumn{1}{c}{539.086$^{***}$} & \multicolumn{1}{c}{285.358$^{***}$} & \multicolumn{1}{c}{158.568$^{***}$} \\ 
%Score (Logrank) Test (df = 12) & \multicolumn{1}{c}{1,906.684$^{***}$} & \multicolumn{1}{c}{1,754.136$^{***}$} & \multicolumn{1}{c}{385.036$^{***}$} & \multicolumn{1}{c}{438.834$^{***}$} & \multicolumn{1}{c}{233.380$^{***}$} & \multicolumn{1}{c}{136.477$^{***}$} \\ 
\bottomrule
\end{tabular}
\begin{tablenotes}
 \footnotesize
 \item \justifying   {\it Notes:} This table reports coefficients of the effect STHC on first-time AI adoption across fields. ***, ** and * indicate significance at the 1\%, 5\% and 10\% level, respectively. The effect of STHC on AI adoption is estimated using a conditional logit with matching. \end{tablenotes}
 \end{threeparttable}
 }
\end{table}

\begin{table}[!h] \centering \centering 
\renewcommand{\arraystretch}{0.8}
\setlength{\tabcolsep}{0.3pt}
  \caption{Conditional logit with matching across fields (re-using AI)} 
  \label{tab:reusingfield} 
\scalebox{0.84}{
\begin{threeparttable}
\begin{tabular}{@{\extracolsep{18pt}}lc c c c c c  } 
\toprule
% & \multicolumn{6}{c}{\textit{Dependent variable: Reusing AI}} \\ 
%\cline{2-7} 
\\[-1.8ex] & \multicolumn{1}{c}{Medicine} & \multicolumn{1}{c}{Biology} & \multicolumn{1}{c}{Chemistry} & \multicolumn{1}{c}{Physics} & \multicolumn{1}{c}{Psychology} & \multicolumn{1}{c}{Materials science} \\ 
\\[-1.8ex] & \multicolumn{1}{c}{(1)} & \multicolumn{1}{c}{(2)} & \multicolumn{1}{c}{(3)} & \multicolumn{1}{c}{(4)} & \multicolumn{1}{c}{(5)} & \multicolumn{1}{c}{(6)}\\ 
\hline \\[-1.8ex] 
\\
\underline{\textit{Realized STHC}} \\
\\
 \hspace{0.5em} \# Domain aut. &  -0.250$^{***}$ & -0.286$^{***}$ & -0.649$^{***}$ & -0.385$^{***}$ & 0.249 & -0.489$^{*}$ \\ 
  & (0.042) & (0.050) & (0.127) & (0.103) & (0.182) & (0.269) \\ 
  & & & & & & \\ 
\hspace{0.5em} \# AI exp. aut. & 0.134$^{***}$ & 0.068 & -0.244$^{**}$ & -0.142 & -0.571$^{***}$ & 0.024 \\ 
  & (0.039) & (0.047) & (0.103) & (0.113) & (0.195) & (0.236) \\ 
  & & & & & & \\ 
\hspace{0.5em} \# CS aut. & -0.094$^{**}$ & -0.138$^{***}$ & -0.041 & -0.019 & 0.309$^{*}$ & -0.146 \\ 
  & (0.040) & (0.045) & (0.103) & (0.104) & (0.182) & (0.232) \\ 
  & & & & & & \\ 
\hspace{0.5em} \# Newbies aut. & 0.229$^{***}$ & 0.101$^{**}$ & 0.263$^{**}$ & 0.336$^{***}$ & -0.444$^{**}$ & 0.649$^{***}$ \\ 
  & (0.036) & (0.044) & (0.103) & (0.105) & (0.182) & (0.246) \\ 
  & & & & & & \\ 
\hspace{0.5em} \# Citations & 0.116$^{***}$ & 0.197$^{***}$ & 0.398$^{***}$ & 0.141$^{**}$ & 0.318$^{***}$ & 0.491$^{***}$ \\ 
  & (0.024) & (0.029) & (0.073) & (0.068) & (0.112) & (0.134) \\ 
  & & & & & & \\ 
\\
\underline{\textit{STHC endowment}} \\
\\  

\hspace{0.5em} AI inst. spe. & 0.027 & 0.095 & 0.240$^{*}$ & 0.010 & 0.520$^{*}$ & 0.185 \\ 
  & (0.090) & (0.082) & (0.132) & (0.129) & (0.295) & (0.222) \\ 
  & & & & & & \\ 
\hspace{0.5em} Inst. cit. & 0.073 & -0.141$^{**}$ & -0.099 & 0.172 & 0.258 & -0.322 \\ 
  & (0.063) & (0.069) & (0.168) & (0.153) & (0.294) & (0.374) \\ 
  & & & & & & \\ 
\hspace{0.5em} Shanghai ranked  & 0.021 & -0.076 & -0.369 & 0.195 & 0.028 & 0.039 \\ 
  & (0.131) & (0.126) & (0.263) & (0.327) & (0.444) & (0.621) \\ 
  & & & & & & \\ 
\hspace{0.5em} HPC & 0.139$^{***}$ & -0.069 & 0.184$^{*}$ & -0.105 & 0.050 & 0.038 \\ 
  & (0.047) & (0.051) & (0.105) & (0.116) & (0.196) & (0.232) \\ 
  & & & & & & \\ 
\hspace{0.5em} \# Domain col. & -0.197$^{***}$ & -0.307$^{***}$ & -0.206 & -0.096 & -0.023 & -0.312 \\ 
  & (0.067) & (0.069) & (0.134) & (0.128) & (0.227) & (0.275) \\ 
  & & & & & & \\ 
\hspace{0.5em} \# CS col.  & 0.057 & 0.183$^{***}$ & 0.103 & -0.121 & -0.163 & -0.189 \\ 
  & (0.045) & (0.049) & (0.092) & (0.106) & (0.182) & (0.202) \\ 
  & & & & & & \\ 
\hspace{0.5em} \# AI col. & 0.292$^{***}$ & 0.184$^{***}$ & 0.291$^{***}$ & 0.368$^{***}$ & 0.627$^{***}$ & 0.418$^{**}$ \\ 
  & (0.039) & (0.042) & (0.091) & (0.092) & (0.170) & (0.178) \\ 
  & & & & & & \\ 
\hspace{0.5em} \# Newbies col.  & 0.224$^{***}$ & 0.295$^{***}$ & 0.119 & -0.013 & 0.069 & 0.168 \\ 
  & (0.049) & (0.053) & (0.103) & (0.097) & (0.180) & (0.213) \\ 
  & & & & & & \\ 
\hspace{0.5em} Exploratory profile & 1.433$^{***}$ & 0.404 & -0.904 & 0.804 & 2.421$^{*}$ & -1.946 \\ 
  & (0.356) & (0.414) & (0.956) & (1.165) & (1.333) & (2.528) \\ 
  & & & & & & \\ 
\hspace{0.5em} Citation stock & 0.017 & -0.032 & -0.078 & 0.045 & 0.050 & 0.093 \\ 
  & (0.020) & (0.025) & (0.053) & (0.049) & (0.085) & (0.117) \\ 
  & & & & & & \\ 
\hspace{0.5em} \% International pub.  & -0.308$^{***}$ & -0.324$^{***}$ & -0.218 & -0.151 & -0.489 & -0.733$^{*}$ \\ 
  & (0.088) & (0.095) & (0.212) & (0.196) & (0.321) & (0.420) \\ 
  & & & & & & \\ 
\hspace{0.5em} Proximity to AI & 0.321$^{***}$ & 0.167$^{***}$ & 0.280$^{**}$ & 0.131 & 0.375$^{***}$ & -0.269 \\ 
  & (0.048) & (0.051) & (0.113) & (0.090) & (0.144) & (0.260) \\ 
  & & & & & & \\ 
\hline \\[-1.8ex] 
Observations & \multicolumn{1}{c}{11,982} & \multicolumn{1}{c}{8,752} & \multicolumn{1}{c}{2,042} & \multicolumn{1}{c}{1,808} & \multicolumn{1}{c}{886} & \multicolumn{1}{c}{520} \\ 
R$^{2}$ & \multicolumn{1}{c}{0.051} & \multicolumn{1}{c}{0.035} & \multicolumn{1}{c}{0.052} & \multicolumn{1}{c}{0.034} & \multicolumn{1}{c}{0.087} & \multicolumn{1}{c}{0.078} \\ 
%Max. Possible R$^{2}$ & \multicolumn{1}{c}{0.500} & \multicolumn{1}{c}{0.500} & \multicolumn{1}{c}{0.500} & \multicolumn{1}{c}{0.500} & \multicolumn{1}{c}{0.500} & \multicolumn{1}{c}{0.500} \\ 
Log Likelihood & \multicolumn{1}{c}{-3,838.484} & \multicolumn{1}{c}{-2,876.278} & \multicolumn{1}{c}{-653.417} & \multicolumn{1}{c}{-595.251} & \multicolumn{1}{c}{-266.871} & \multicolumn{1}{c}{-159.214} \\ 
%Wald Test (df = 17) & \multicolumn{1}{c}{541.100$^{***}$} & \multicolumn{1}{c}{283.800$^{***}$} & \multicolumn{1}{c}{92.900$^{***}$} & \multicolumn{1}{c}{56.610$^{***}$} & \multicolumn{1}{c}{62.490$^{***}$} & \multicolumn{1}{c}{33.920$^{***}$} \\ 
LR Test & \multicolumn{1}{c}{628.321$^{***}$} & \multicolumn{1}{c}{313.869$^{***}$} & \multicolumn{1}{c}{108.573$^{***}$} & \multicolumn{1}{c}{62.708$^{***}$} & \multicolumn{1}{c}{80.386$^{***}$} & \multicolumn{1}{c}{42.009$^{***}$} \\ 
%Score (Logrank) Test (df = 17) & \multicolumn{1}{c}{596.553$^{***}$} & \multicolumn{1}{c}{303.224$^{***}$} & \multicolumn{1}{c}{102.896$^{***}$} & \multicolumn{1}{c}{60.548$^{***}$} & \multicolumn{1}{c}{73.705$^{***}$} & \multicolumn{1}{c}{38.981$^{***}$} \\ 
\bottomrule 
\end{tabular} 
\begin{tablenotes}
 \footnotesize
 \item \justifying   {\it Notes:} This table reports coefficients of the effect STHC on re-using AI across fields. ***, ** and * indicate significance at the 1\%, 5\% and 10\% level, respectively. The effect of STHC on re-using AI is estimated using a conditional logit with matching. \end{tablenotes}
 \end{threeparttable}
 }
\end{table}

\end{document}